\documentclass{elsarticle}
\usepackage{graphicx} 
\usepackage{amsmath}
\usepackage{amssymb}
\usepackage{amsfonts}
\usepackage{natbib}
\usepackage{IEEEtrantools}
\usepackage{appendix}
\newcommand{\ai}{\emph{ab initio}}
\newcommand{\EinA}{Einstein \emph{A} coefficients}
\newcommand{\etal}{\emph{et al.}}
\usepackage{color}

\newcommand{\cm}{cm$^{-1}$}
\biboptions{sort&compress}

\title{Effect of the avoided crossing on the rovibrational energy levels, resonances, and predissociation lifetimes within the ground and first excited electronic states of lithium fluoride}

\date{November 9, 2025}

\author{V.~G.~Ushakov$^a$, A.~Yu.~Ermilov$^b$, E.~S.~Medvedev$^a$}

\address{$^a$Federal Research Center of Problems of Chemical Physics and Medicinal Chemistry (former Institute of Problems of Chemical Physics), Russian Academy of Sciences, 142432 Chernogolovka, Russian Federation}
\address{$^b$M.~V.~Lomonosov Moscow State University, Russian Federation}

\begin{document}

\begin{abstract}
We investigate the LiF spectrum up to $7800$ cm$^{-1}$ above the first dissociation limit. The \ai\ calculations of the adiabatic potentials and other molecular functions are performed in a wide range of interatomic separations, $r=1$-17 bohr.
We consider the model of two interacting electronic states including both the bound states and the resonances of two kinds, the tunneling resonances and the predissociative ones.
The Born-Oppenheimer potentials are modeled with use of two functions imitating the diabatic terms whereas the coupling between them was set constant
equal to the half of the minimum separation between the adiabatic terms, and then we define the final diabatic terms and the final diabatic coupling \emph{via} the adiabatic potentials and the angle of the adiabatic-to-diabatic basis rotation obtained by integration of the nonadiabatic coupling matrix element. The energies of the bound states, as well as the positions and widths of the resonances are calculated. The observed transition frequencies are reproduced with the standard deviation of 0.0009 cm$^{-1}$ for $^7$LiF, 0.0006 cm$^{-1}$ for $^6$LiF, and within the experimental uncertainties for the most of the lines.
The line lists for the bound-bound $X$-$X$ rovibrational transitions are calculated for quantum numbers $v\le50,\Delta v\le15,J\le170$ ($J\le200$ for the 0-0 and 1-0 bands).

\end{abstract}

\maketitle

\section{Introduction}\label{intro}

The avoided crossing between two lowest $1^1\Sigma^+$ ($X$) and $2^1\Sigma^+$ ($B$) electronic terms of LiF and other alcali halides is a well-known phenomenon in the theory of gas-phase reactive atomic collisions \cite{Grice74} and collisional dissociation \cite{Ewing71}. The objective of those studies was characterization of the nonadiabatic chemi-ionization processes M+X$\xrightarrow{}$M$^+$+X$^-$ \cite{Kahn74,Werner81} and photodissociation \cite{Valiev20}. However, in calculations of the transitions between the rovibratonal levels of the $X$ state in frames of the one-state model \cite{Bittner18}, this phenomenon was ignored because the curve crossing occurs at about 13.7 bohr, far away from equilibrium at $\approx3$ bohr. Strictly speaking, this is not justified even for low $v$ and $\Delta v$ transitions if one needs a very high, spectroscopic, accuracy 
but it definitely needs to be taken into account for higher levels and overtones especially in a vicinity of the crossing.

The focus of the previous theoretical studies \cite{Kahn74,Werner81,Bauschlicher88,Giese04,Varandas09,Nkambule15,Bittner18,Valiev20} was on the \ai\ calculations of the \emph{adiabatic} potential-energy functions (PEFs), the \emph{nonadiabatic} coupling matrix element (NACME) and the dipole-moment functions (DMFs). However, numerically solving the Schr\"odinger equation in the adiabatic representation encountered difficulties caused by rapid variations of the adiabatic PEFs and NACME in a vicinity of the avoided crossing. Therefore, a diabatization procedure was introduced  \cite{Smith69,Baer75,Delos79}  in order to remove the dynamic coupling due to the first derivative of the electronic wavefunction over the internuclear distance, $r$. The unitary matrix that performs the rotation from the adiabatic basis to the diabatic one, and the rotation angle, $g(r)$, are expressed in terms of the non-adiabatic-coupling matrix element (NACME).
In particular, Varandas \cite{Varandas09} found the rotation matrix with use of the \ai\ calculations of the NACME. He interpolated the NACME with several model functions, and found, in particular, that the Lorentzian curve decayed too slowly at the wings.

The diabatic terms and the coupling were
expected to behave smoothly. In particular, Smith \cite{Smith69} assumed and showed that, in a good approximation, the smooth diabatic terms can be obtained by neglecting interactions between the ionic and covalent configurations.

Instead of direct calculation of the rotation matrix, many authors proposed simplified diabatization schemes based on the physically grounded models of the ionic and covalent diabatic terms. Thus, 
Kahn \etal\ \cite{Kahn74} and Bauschlicher\& Langhoff \cite{Bauschlicher88} did not calculate NACME and $g(r)$, rather they defined the diabatic ionic potential by matching the Rittner function \cite{Rittner51} with the adibatic ground-state potential at $r=6.2$ bohr.
Werner and Meyer \cite{Werner81} 
defined the rotation angle by diagonalizing the matrix of the dipole moments. They showed that the rotation angle thus calculated well reproduced the behavior, in a vicinity of the avoided crossing, of the angle directly calculated \emph{via} NACME. 

The dibatization procedures described above resulted in a rather complicated behavior of the diabatic coupling as function of $r$ shown, \emph{e.g.}, in Fig. 1 of Erratum to Ref. \cite{Varandas09} along with the results of other authors. 
These results have never been used to calculate any observables like the reaction cross section or the rovibrational transition probability.

In contrast, Child \cite{Child69} and other authors \cite{Voronin83,Volokhov85} calculated the cross section for the reactive scattering in frames of the simple Landau-Zener model where the diabatic coupling was set constant. This is physically justified because the coupling is expected to vary slowly within a narrow vicinity of the crossing and because its behavior outside this region is of no importance due to increasing separation of the adiabatic terms.

In this paper, we investigate the LiF spectrum up to $E=7800$ cm$^{-1}$ above the first dissociation limit, $E=0$, approximately half way to the second dissociation limit. 
We consider the model of two interacting electronic states including both the bound states and the resonances of two kinds, the tunneling resonances and the predissociative ones.
After formulating the system of two coupled adiabatic equations for the $X$ and $B$ states in Sec. \ref{twostates} and description of the \ai\ calculations in Sec. \ref{aicalc}, we introduce the model functions for the PEFs and NACME in Sec. \ref{molfunc}. The Born-Oppenheimer PEFs are modeled with use of two functions imitating the dibatic terms whereas the coupling between them was set constant
equal to the half of the minimum separation between the adiabatic terms. The NACME is modeled with a function that decays at the wings faster than the Lorentzian. The parameters of the model molecular functions were optimized by fitting to the \ai\ calculated PEFs, NACME, and the observed transitions frequencies. Then, we define the final diabatic terms and the final diabatic coupling \emph{via} the adiabatic PEFs and $g(r)$ obtained by integration of the model NACME. The energies of the bound states, the positions and widths of the resonances, and the line lists for the $X$-$X$ transitions in two isotopologues are calculated. The observed transition frequencies are reproduced with the spectroscopic accuracy of $\sim0.001$ cm$^{-1}$.

\section{The two-states model}\label{twostates}

The full wavefunction in the adiabatic (clamped-nuclei) basis has the form
\begin{equation}
    \Psi(q;r) = \phi_1^{\textrm{ad}}(q;r)\varphi_1^{\textrm{ad}}(r)+\phi_2^{\textrm{ad}}(q;r)\varphi_2^{\textrm{ad}}(r), \label{Psi_ad}
\end{equation}
where $q$ is a set of electronic coordinates, $r$ the internuclear separation, 1 and 2 refer to the $X$ and $B$ states, respectively, $\phi_{1,2}^{\textrm{ad}}$ are the adiabatic wavefunctions\footnote{Here and below, the arguments of functions are omitted after their definition.} of the 
electronic Born-Oppenheimer (BO) states with energies $U_{1,2}^{\textrm{BO}}(r)$, and $\varphi_{1,2}^{\textrm{ad}}$ are the adiabatic nuclear amplitudes.
The nonadiabatic corrections to the potentials, $\mathcal{N}_{ik}(r)$, and the NACME between the adiabatic states, $\mathcal{N}(r)$, are given by
\begin{IEEEeqnarray}{ccl}
    \mathcal{N}_{ik}(r) &=& -\frac{\hbar^2}{2\mu}\left<\phi_i^\textrm{ad}\frac{\partial^2}{\partial r^2}\phi_k^\textrm{ad}\right>,\textrm{          }i,k=1,2, \label{nacme2}\\
    \mathcal{N}(r) &=& \left<\phi_1^{\textrm{ad}}\frac{\partial}{\partial r}\phi_2^{\textrm{ad}}\right>=-\left<\phi_2^{\textrm{ad}}\frac{\partial}{\partial r}\phi_1^{\textrm{ad}}\right>, \label{nacme1}
\end{IEEEeqnarray}
where $\left<...\right>$ stands for integration over $q$; the second equality in Eq. (\ref{nacme1}) follows from the orthogonality condition, $\left<\phi_1^{\textrm{ad}}\phi_2^{\textrm{ad}}\right>\equiv0$ at all $r$. The mass-dependent corrections and the NACME are small.
Nevertheless, they can significantly affect the energy levels, resonances, and transition probabilities at high-enough energies.

The adiabatic nuclear amplitudes, $\varphi_{1,2}^{\textrm{ad}}$, are solutions of the system of two coupled Schr\"odinger equations,
\begin{equation} \label{eqsAD}
    \left[-{\frac{\hbar^2}{2\mu}\frac{d^2}{dr^2}}+U_J(r)+\left(\begin{tabular}{cc}
       $U_1^\textrm{ad}$ & $\mathcal{N}_{12}-\frac{\hbar^2}{\mu}\mathcal{N}\frac{d}{dr}$\\
       $\mathcal{N}_{21}+\frac{\hbar^2}{\mu}\mathcal{N}\frac{d}{dr}$  &  $U_2^\textrm{ad}$ 
    \end{tabular} \right) -E \right] \left( \begin{tabular}{c}
         $\varphi_1^{\textrm{ad}}$ \\
         $\varphi_2^{\textrm{ad}}$ 
    \end{tabular}\right) = 0,
\end{equation}
where $\mu$ is the reduced mass of two atoms, $U_J=\hbar^2J(J+1)/2\mu r^2$ is the rotational energy, and the adiabatic potentials include the mass-dependent nonadiabatic contributions,
\begin{IEEEeqnarray}{ccl}
    U_1^\textrm{ad} &=& U_1^{\textrm{BO}} + \mathcal{N}_{11}, \label{U1BO}\\
    U_2^\textrm{ad} &=& U_2^{\textrm{BO}} + \mathcal{N}_{22}. \label{U2BO}
\end{IEEEeqnarray}
The adiabatic electronic wavefunctions, $\phi_{1,2}^\textrm{ad}$, the BO potentials, and the NACME of Eq. (\ref{nacme1}) are calculated \ai\ in a wide range of $r$ using the MOLPRO package \cite{MOLPRO2010}. 

Note that, in general, the system of two equations (\ref{eqsAD}) is not Hermitian because $\mathcal{N}$ and $d/dr$ do not commute. Yet 
it is easy to verify by differentiating over $r$ twice the above-mentioned orthogonality condition that the nonadiabatic corrections must satisfy the following relation:
\begin{equation}\label{Hermitian}
    \mathcal{N}_{21}-\mathcal{N}_{12} = \frac{\hbar^2}{\mu}\frac{d\mathcal{N}}{dr},
\end{equation}
which guarantees the system being Hermitian. 

The above-introduced adiabatic potentials and the nonadiabatic contributions 
are all point-wise since they are obtained as a result of \ai\ calculations, hence they should be interpolated for use in Eqs. (\ref{eqsAD}). The interpolations will be performed in Sec. \ref{molfunc}.

For calculations of the transition dipole moment (TDM), we need three matrix elements of the dipole-moment operator,
\begin{IEEEeqnarray}{ccc}
    d^\textrm{ad}_X(r) &=& \left<\phi_1^{\textrm{ad}}\hat{d}_z\phi_1^{\textrm{ad}}\right>,\label{dX}\\
    d^\textrm{ad}_B(r) &=& \left<\phi_2^{\textrm{ad}}\hat{d}_z\phi_2^{\textrm{ad}}\right>,\label{dA}\\
    d^\textrm{ad}_{XB}(r) &=& \left<\phi_1^{\textrm{ad}}\hat{d}_z\phi_2^{\textrm{ad}}\right>.\label{dXA}
\end{IEEEeqnarray}
The TDMs between the upper rovibrational state (primed) and the lower state (double primed) are non zero for $J^\prime-J^{\prime\prime}=-1$ (the P branch) or $+1$ (the R branch). In both cases, they have the form
\begin{equation}\label{TDM}
    \textrm{TDM} = \sqrt{ \frac{J^2-M^2}{4J^2-1}}d,
\end{equation}
where $J=\max(J^\prime,J^{\prime\prime})$, $M$ is projection of the angular momentum onto the $z$ axis of the laboratory coordinate system (LCS), and
\begin{equation}\label{d_ad}
    d=d_{11}^\textrm{ad}+d_{22}^\textrm{ad}+d_{12}^\textrm{ad}+d_{21}^\textrm{ad}.
\end{equation}
Parameters $d_{ik}^\textrm{ad}$ are integrals of the DMFs of Eqs. (\ref{dX})-(\ref{dXA}) with the adiabatic nuclear amplitudes,
\begin{IEEEeqnarray}{ccl}
    d_{11}^\textrm{ad} &=&\int{\varphi_1^{\textrm{ad}}}^\prime d^\textrm{ad}_X {\varphi_1^{\textrm{ad}}}^{\prime\prime}dr,\label{d11} \\
    d_{22}^\textrm{ad} &=&\int{\varphi_2^{\textrm{ad}}}^\prime d^\textrm{ad}_B {\varphi_2^{\textrm{ad}}}^{\prime\prime}dr,\label{d22}\\
    d_{ik}^\textrm{ad} &=&\int{\varphi_i^{\textrm{ad}}}^\prime d^\textrm{ad}_{XB} {\varphi_k^{\textrm{ad}}}^{\prime\prime}dr,\label{dik} \hspace{10pt} i\ne k.
\end{IEEEeqnarray}

The system of two equations (\ref{eqsAD}) is a part of an infinite system of coupled equations. The nonadiabatic couplings between them are generally small and rapidly decrease with increasing the gap between the adiabatic terms. Thus, as a rule, the one-state model applies, as was done by Bittner and Bernath \cite{Bittner18}. However, the couplings strongly increase near the avoided crossing, hence, necessity arises to consider the two-states model. 

The BO potentials, the nondiabatic couplings, and the adiabatic wavefunctions change very rapidly in a vicinity of the avoided crossing, which creates difficulties in numerical calculations. Therefore, the so-called diabatization procedure has been developed \cite{Smith69,Baer75,Kahn74,Delos79,Werner81,Bauschlicher88,Varandas09,MOLPRO2010} that removes the dynamic coupling due to the $d/dr$ operator in Eqs. (\ref{eqsAD}). The procedure consists in a transformation of the adiabatic electronic wavefunctions and nuclear amplitudes to the diabatic ones, $\phi_{1,2}^{\textrm{d}}(q;r)$ and $\varphi_{1,2}^{\textrm{d}}(r)$, using the unitary matrix of rotation by angle $g(r)$,
\begin{equation}\label{d2ad}
    \phi^\textrm{d} = \textbf{u}\phi^\textrm{ad}\textrm{     and     }\varphi^\textrm{d} = \textbf{u}\varphi^\textrm{ad},
\end{equation}
where \cite{Werner81}
\begin{equation}\label{umatr}
    \textbf{u} = \left( \begin{tabular}{cc}
    $\cos{g}$    &  $\sin{g}$\\
    $-\sin{g}$   & $\cos{g}$
    \end{tabular} \right),\hspace{5pt}\phi^\textrm{d} = \left( \begin{tabular}{c}
         $\phi_1^\textrm{d}$  \\
         $\phi_2^\textrm{d}$
    \end{tabular} \right), \hspace{5pt}\emph{etc.}
\end{equation}
Angle $g$ is selected such that the dynamic coupling disappears \cite{Smith69},
\begin{equation}\label{Nd=0}
        \left<\phi_1^{\textrm{d}}\frac{\partial}{\partial r}\phi_2^{\textrm{d}}\right>=0. 
\end{equation}
The condition for this is \cite{Smith69,Baer75,Delos79,Werner81}
\begin{equation}\label{g}
    \frac{dg}{dr} = \mathcal{N}.
\end{equation}
For the purpose of integration to obtain $g$, the \ai\ $\mathcal{N}$ values will be interpolated in Sec. \ref{molfunc}. 

The diabatic nuclear amplitudes are obtained as solutions of the system of equations,
\begin{equation} \label{eqsDiab}
    \left[-{\frac{\hbar^2}{2\mu}\frac{d^2}{dr^2}}+U_J+\left(\begin{tabular}{cc}
       $U_1^{\textrm{d}}$     &  $V^{\textrm{d}}$\\
       $V^{\textrm{d}}$  &  $U_2^{\textrm{d}}$
    \end{tabular} \right)-\frac{\hbar^2}{2\mu}\mathcal{N}^2-E \right] \left( \begin{tabular}{c}
         $\varphi_1^{\textrm{d}}$ \\
         $\varphi_2^{\textrm{d}}$ 
    \end{tabular}\right) = 0,
\end{equation}
where the diabatic potentials, $U_{1,2}^\textrm{d}(r)$, and the (static) diabatic coupling, $V^\textrm{d}(r)$, are obtained from the adiabatic potentials and the rotation angle as
\begin{IEEEeqnarray}{lcl}
    U_1^\textrm{d} &=& U_1^\textrm{ad}\cos^2{g}+U_2^\textrm{ad}\sin^2{g}+\mathcal{M}\sin{2g},\label{U1d}\\
    U_2^\textrm{d} &=& U_1^\textrm{ad}\sin^2{g}+U_2^\textrm{ad}\cos^2{g}-\mathcal{M}\sin{2g},\label{U2d}\\
    V^\textrm{d} &=& \tfrac{1}{2}\left(U_2^\textrm{ad}-U_1^\textrm{ad} \right)\sin{2g}+\mathcal{M}\cos{2g},\label{Vd}\\
    \mathcal{M}&=&\tfrac{1}{2}\left(\mathcal{N}_{12}+\mathcal{N}_{21}\right). \label{M}
\end{IEEEeqnarray}
The latter equations can be divided into the BO and non-BO parts,
\begin{IEEEeqnarray}{ccl}
    U_{1,\textrm{BO}}^\textrm{d} &=& U_1^\textrm{BO}\cos^2{g}+U_2^\textrm{BO}\sin^2{g},\label{U1dBO}\\
    U_{2,\textrm{BO}}^\textrm{d} &=& U_1^\textrm{BO}\sin^2{g}+U_2^\textrm{BO}\cos^2{g},\label{U2dBO}\\
    V_{\textrm{BO}}^\textrm{d} &=& \tfrac{1}{2}\left(U_2^\textrm{BO}-U_1^\textrm{BO} \right)\sin{2g},\label{VdBO}\\
    U_{1,\textrm{NBO}}^\textrm{d} &=& \mathcal{N}_{11}\cos^2{g}+\mathcal{N}_{22}\sin^2{g} + \mathcal{M}\sin{2g},\label{U1dNBO}\\
    U_{2,\textrm{NBO}}^\textrm{d} &=& \mathcal{N}_{11}\sin^2{g}+\mathcal{N}_{22}\cos^2{g} - \mathcal{M}\sin{2g},\label{U2dNBO}\\
    V_{\textrm{NBO}}^\textrm{d} &=& \tfrac{1}{2}\left(\mathcal{N}_{22}-\mathcal{N}_{11} \right) \sin{2g} + \mathcal{M}\cos{2g},\label{VdNBO}
\end{IEEEeqnarray}

Equations (\ref{eqsDiab}) are simpler to solve because the potentials and the coupling are smooth functions everywhere including the domain of the avoided crossing. Combining their solutions for the diabatic nuclear amplitudes, $\varphi^\textrm{d}$, with the electronic diabatic wavefunctions, $\phi^\textrm{d}$, of Eq. (\ref{d2ad}), we can rewrite the full wavefunction of Eq. (\ref{Psi_ad}) as
\begin{equation}
    \Psi = \phi_1^{\textrm{d}}\varphi_1^{\textrm{d}}+\phi_2^{\textrm{d}}\varphi_2^{\textrm{d}}. \label{Psi_d}
\end{equation}

The diabatic electronic wavefunctions are not eigenfunctions of the electronic Hamiltonian \cite{Werner81} since they are linear combinations of the adiabatic wavefunctions of the states with different energies, $U_1^\textrm{BO}$ and $U_2^\textrm{BO}$. They can be interpreted as an incomplete basis set of functions containing, \emph{e.g.}, only ionic or only covalent configurations. Thus, they have no physical meaning and should be considered only as a useful tool in calculations of the Li + F $\rightarrow$ Li$^+$ + F$^-$ chemi-ionization process \cite{Kahn74,Werner81}, LiF photodissociation dynamics \cite{Balakrishnan99}, or LiF
$X$-$X$ rovibrational and $X$-$B$ rovibronic transition intensities in a complicated situation of the avoided crossing. The calculations of the intensities can be done with use of the same equations (\ref{dX})-(\ref{dik}) in which all adiabatic functions and parameters are replaced with the diabatic ones and the diabatic dipole-moment functions are expressed in terms of the adiabatic ones by the relations
\begin{eqnarray}
d^\textrm{d}_X &=&d^\textrm{ad}_X\cos ^{2}g+d^\textrm{ad}_B\sin ^{2}g+d_{XB}^\textrm{ad}\sin 2g,\label{ddiabaticX} \\
d^\textrm{d}_B &=&d^\textrm{ad}_X\sin ^{2}g+d^\textrm{ad}_B\cos ^{2}g-d^\textrm{ad}_{XB}\sin 2g, \label{ddiabaticB}\\
d^\textrm{d}_{XB} &=&\frac{1}{2}\left( d^\textrm{ad}_B-d^\textrm{ad}_X\right)\sin
2g+d^\textrm{ad}_{XB}\cos 2g. \label{ddiabaticXB}
\end{eqnarray}%

The calculation procedure to be developed further 
is briefly summarized as follows. First, the \ai\ calculations are performed in Sec. \ref{aicalc} to obtain all necessary adiabatic functions entering Eqs. (\ref{nacme2})-(\ref{U2BO}). Second, the transformation to the diabatic basis by means of the rotation matrix is introduced.
Third, the analytic functions for modeling the \ai\ functions are constructed in Sec. \ref{molfunc}, and their parameters are optimized by means of least-squares fitting to the \ai\ and experimental data; in the fitting process, the Schr\"odinger equation in the diabatic basis, Eq. (\ref{eqsDiab}), is solved by the adaptive sinc-DVR method \cite{Colbert92sincDVR,Meshkov08}. The final solution with the optimized parameters provides for positions of the energy levels, the resonances,
and the predissociation lifetimes of the quasi-bound states.


\section{\emph{Ab initio} calculations}\label{aicalc}  

	The Born-Oppenheimer potentials of the \emph{X} and \emph{B} states, the dipole-moment matrix elements, the mass-dependent corrections to the BO PEFs, and the NACME for LiF molecule are calculated at the MRDCI level by the MOLPRO-2010.1 program package \cite{MOLPRO2010}. The calculations are carried out for wide internuclear distances within 1-17 bohr. The d-aug-cc-pv5z basis set with double set diffuse functions is chosen at the F atom for the better description of the fluorine anion; the aug-cc-pv5z is chosen at Li. The active space included 11 MO and 8 active electrons. The orbitals are optimized using the state-averaged CASSCF procedure with equal weights for the ionic and neutral solutions. The total numbers of uncontracted configurations in the MRDCI was about 214 millions. The corrections to the potentials and NACME are calculated by the numerical differentiation using the DDR procedure of the MOLPRO package.

\section{Modeling the molecular functions}\label{molfunc} 

The molecular functions can be modeled in both the adiabatic and diabatic bases, the choice being dictated by the known properties of the functions.

Interpolations of the BO potentials will be performed in the adiabatic basis because we know how their rapid variations near the avoided crossing arises from smooth behavior of the diabatic terms. The model functions are selected in the form
\begin{equation}
U_{1,2}^{\,\mathrm{BO}}=\tfrac{1}{2} \left(u_{1}+u_{2}\right)\mp \tfrac{1}{2} \sqrt{\left( 
u_{2}-u_{1}\right)^{2}+4V^{2}} , \label{U12BO}
\end{equation}
where $V$ is constant equal to half of the minimum separation between the adiabatic terms at the avoided crossing, $r_\textrm{c}$; ``minus'' and ``plus'' relate to the ground and excited states, respectively. Our calculations for LiF gave $r_\textrm{c}=13.6902$ bohr and $V=139.7$ cm$^{-1}$. Functions $u_{1,2}$ and constant $V$ imitate smooth diabatic potentials and constant diabatic coupling between them; they can be considered as a first approximation to the true diabatic potentials and coupling given by Eqs. (\ref{U1d})-(\ref{Vd}).

The analytic forms of $u_1$ and $u_2$ are similar to those of Meshkov previously used in the CO studies \cite{Meshkov18},
\begin{IEEEeqnarray}{ccl}
    u_1(r) &=& Y_1(r)S_1(r)+\Delta, \label{Y1}\\
    u_2(r) &=& \left(Y_2(r)+\delta\right)S_2(r)-\delta, \label{Y2}
\end{IEEEeqnarray}
where $\Delta=16066.56$ cm$^{-1}$ is the asymptotic difference of the energies in the two states ($u_2(\infty)=0$ and the asymptotic energy of the adibatic ground-state term is approximately zero) and $\delta=100$ cm$^{-1}$ is introduced in order to avoid the 0/0 uncertainty at $r\rightarrow\infty$.

Functions $Y_{1,2}$ are further decomposed into the short- and long-range parts,
\begin{IEEEeqnarray}{ccl}
    Y_1(r) &=& Y_0^X(r)+Y_1^\textrm{LR}(r), \label{Y01LR}\\
    Y_2(r) &=& Y_0^B(r).
\end{IEEEeqnarray}
Functions $Y_0^{X,B}$ describe the behavior of the potentials in the united-atom limit,
\begin{equation}\label{Y0}
    Y_0^X = \left[\frac{K}{r}+E_0+K\alpha+\left(E_0+\frac{1}{2}K\alpha\right)\alpha r\right]e^{-\alpha r},
\end{equation}
where $K = 3.135806278\cdot10^6$  cm$^{-1}$\AA, $E_0 = -2.041957\cdot10^7$ cm$^{-1}$, $\alpha=7$ \AA$^{-1}$, and
\begin{equation}
Y_0^B=\frac{K}{r}e^{-\beta r}+\left[ \left( E_{0}+K\beta \right)
\left( 1+\gamma r\right) -\frac{1}{2}K\beta ^{2}r\right] e^{-\gamma
r},
\end{equation}
where $\beta=3.6$ \AA$^{-1}$, $\gamma=10$ \AA$^{-1}$. At $\alpha = \beta = \gamma$, functions
$Y_0^X$ and $Y_0^B$ are identical.
The long-range part of $u_1$ is the the Rittner potential with minus sign and with an additional factor removing the singularity at the origin,
\begin{equation}\label{Y1LR}
    Y_1^\textrm{LR} = \left(\frac{116140.9733}{r}+\frac{117830.4323}{r^4}\right)\left(1-e^{-\alpha r}\right)^6 \textrm{  cm$^{-1}$}.
\end{equation} 
Dimensionless function $S_1$ is represented in the form
\begin{equation}\label{S1}
    S_1 = \sum_{i=0}^6c_iT_i\left(z_1\right) + \sum_{i=1}^6d_iT_i\left(z_2\right),
\end{equation}
where $T_i(z)$ is Chebyshev polynomial and
\begin{IEEEeqnarray}{lcl}
    z_k = \tanh{\left(a_kr-\frac{b_k}{r}\right)},\hspace{10pt}k=1,2, \nonumber\\
    c_0 = -c_2-c_4-c_6-d_2-d_4-d_6, \nonumber\\
    c_1=-1-c_3-c_5-d_1-d_3-d_5 \nonumber.
\end{IEEEeqnarray}
The latter restrictions on the coefficients result from the boundary conditions $S_1(0)=1, S_1(\infty)=-1$. Similarly,
\begin{equation}\label{S2}
    S_2 = \sum_{i=0}^6\gamma_iT_i\left(\zeta_1\right)+\sum_{i=1}^6\delta_iT_i\left(\zeta_2\right),
\end{equation}
where
\begin{IEEEeqnarray}{lcl}
    \zeta_k = \tanh{\left(\alpha_kr-\frac{\beta_k}{r}\right)},\hspace{10pt}k=1,2, \nonumber\\
    \gamma_0 = 1-\gamma_2-\gamma_4-\gamma_6-\delta_2-\delta_4-\delta_6, \nonumber\\
    \gamma_1=-\gamma_3-\gamma_5-\delta_1-\delta_3-\delta_5 \nonumber,    
\end{IEEEeqnarray}
with the boundary conditions $S_2(0)=S_2(\infty)=1$.

The variable parameters are $a_k,b_k,c_i,d_i,\alpha_k,\beta_k,\gamma_i,\delta_i$, their initial values are obtained by fitting Eqs. (\ref{U12BO}) to the \ai\ data.

\begin{figure}[htbp]
    \centering
    \includegraphics[scale=0.3]{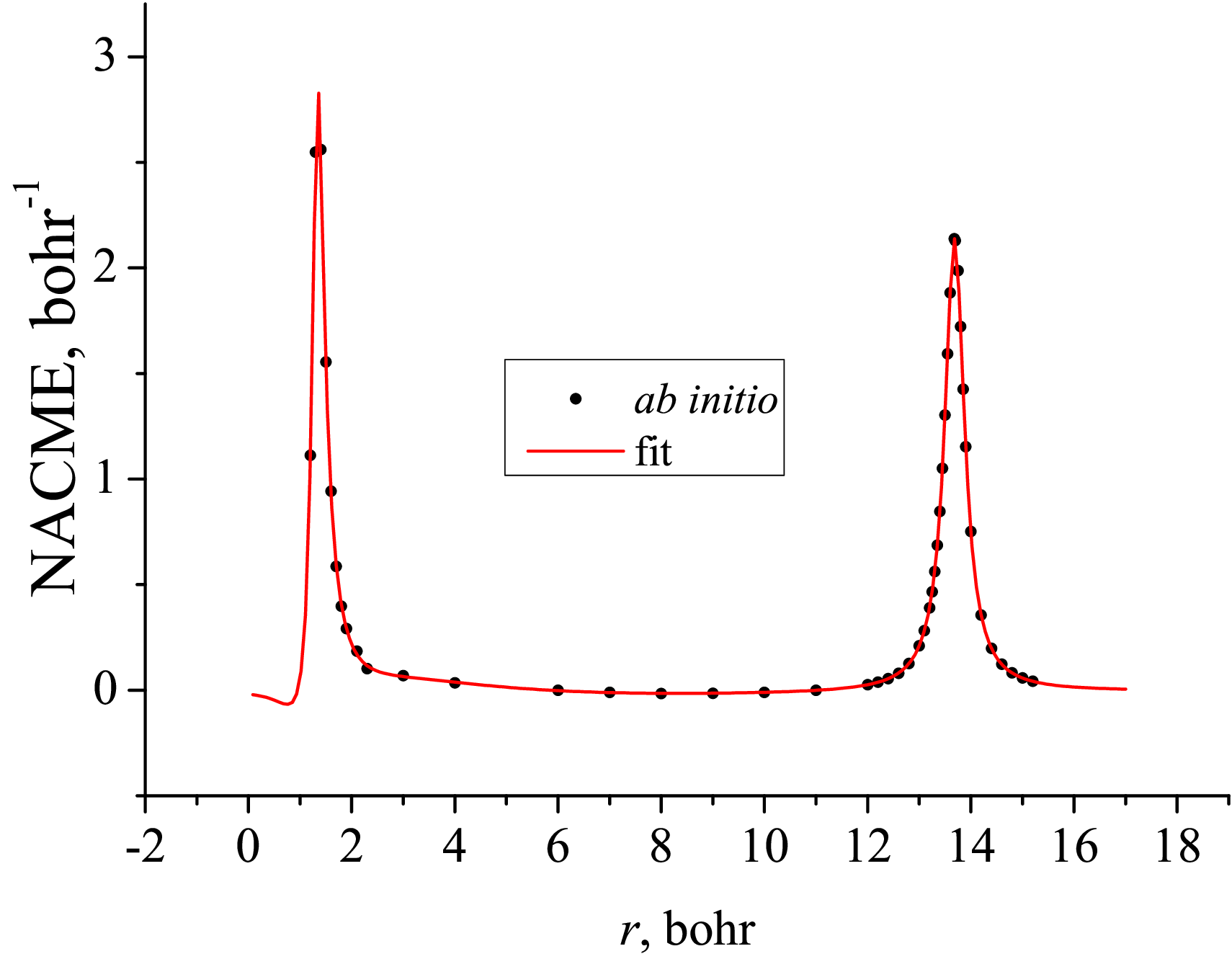}
    \caption{The \ai\ values of $\mathcal{N}$ (points) and the fit.}
    \label{fig:NACME_N}
\end{figure}

Interpolation of NACME of Eq. (\ref{nacme1}) requires special consideration. 
Varandas \cite{Varandas09} found that the Loretzian function interpolated the NACME in a narrow vicinity of $r_\textrm{c}$ very well. However, we found (in accord with Varandas) that the Lorentzian, when moving further away from its maximum, decays too slowly; moreover, there is a second avoided crossing at short $r$ as seen in Fig. \ref{fig:NACME_N}. Therefore, the following function was constructed:
\begin{eqnarray}
\mathcal{N}(r) &=&r^{2}\left\{ \frac{A_{1}}{\left[ \left(
r^{2}-r_{1}^{2}\right) ^{2}+w_{1}^{2}\right] }+\frac{A_{2}}{\left[
\left(
r^{2}-r_{2}^{2}\right) ^{2}+w_{2}^{2}\right] }\right\} S(r), \\
S &=&1+\sum_{i=1}^{6}n_{i}T_i(y),\quad y=\tanh \left(
pr-\frac{q}{r}\right) .
\end{eqnarray}
Variable parameters $p,q,A_{1},r_{1},w_{1},A_{2},r_{2},w_{2},n_{i}$ are optimized to reproduce the \ai\ data. 

As shown in Fig. \ref{fig:NACME_N}, this function excellently reproduces the \ai\ data so that we can use it to obtain $g(r)$ from Eq. (\ref{g}).  

Among other nonadiabatic contributions, only $\mathcal{N}_{11}$ is taken into account since the experimental information is insufficient to determine the others. It is presented in the form used for CO \cite{Meshkov18},
\begin{equation}
\mathcal{N}_{11}=\frac{\hbar^2}{\mu}\frac{a_0+a_1z+a_2z^2+a_3z^3}{1+r^6},
\end{equation}
where
\begin{equation}
z=\tanh \left(\frac{r}{R_0}-\frac{R_0}{r}\right), \quad
R_0=1.56\mathrm{\ \mathring{A}} .
\end{equation}

The mass-dependent $\mathcal{N}_{11}$, while being small, is very important for reaching the experimental accuracy since it permits to describe two isotopologues simultaneously. Similarly working is $\mathcal{N}^2/2$ at the diagonal of the diabatic equations; $\mathcal{N}_{22}$ and $\mathcal{M}$ are probably important for the upper levels as well but no necessary experimental information to determine them is available.

The rotation angle to the diabatic basis is calculated with use of the interpolated $\mathcal{N}$ by equation 
\begin{equation}
    g=\frac{\pi}{4}+\int_{r_0}^r\mathcal{N}dr,
\end{equation}
where $r_0=13.662$ bohr = 7.229616 \AA\ is the coordinate at which $d_X^\textrm{ad}=d_B^\textrm{ad}$. It should be noted that angle $g$ is defined up to an arbitrary constant, and the form of the molecular functions depends on this constant. The above choice of $r_0$ corresponds to the simplest behavior of the off-diagonal DMF in the diabatic basis, $d_{XB}^\textrm{d}$, see Eq. (\ref{ddiabaticXB}). Namely, it vanishes at $r_0$ and remains small in a rather wide vicinity of this point.

Inserting the adiabatic BO potentials of Eqs. (\ref{U12BO}) into Eqs. (\ref{U1dBO})-(\ref{VdBO}), we will express the diabatic BO functions in terms of $u_{1,2}$ as
\begin{IEEEeqnarray}{ccc}
    U_{1,\textrm{BO}}^\textrm{d} &=& u_+ - u_-\cos{2g},\\
    U_{2,\textrm{BO}}^\textrm{d} &=& u_+ + u_-\cos{2g},\\
    V_\textrm{BO}^\textrm{d} &=& u_-\sin{2g},\label{VdBOuu}\\
    u_+ &=& \tfrac{1}{2}\left(u_2+u_1\right),\\
    u_- &=& \tfrac{1}{2}\sqrt{\left(u_2-u_1\right)^2+4V^2}
\end{IEEEeqnarray}

In contrast to the potentials and other above-mentioned functions, the DMFs will be modeled in the diabatic basis. The reason is that the behavior of the DMFs in the ionic and covalent diabatic states is physically clear.
The DMFs in the diabatic basis, Eqs. (\ref{ddiabaticX})-(\ref{ddiabaticXB}), are interpolated with the following functions. The DMF of the diabatic ionic term is
\begin{eqnarray}
d^\textrm{d}_X(r) &=&\left( y_{0}-a_{1}r-\frac{a_{2}}{r^{2}}\right)
\left(
1-e^{-\alpha r}\right) ^{5}\sum_{i=0}^{6}\beta _{i}T_i(y),\label{ddX} \\
y &=&\tanh (\beta r). 
\end{eqnarray}
Parameters $\beta_i$ are restricted by the conditions that the sum in the right-hand side of Eq. (\ref{ddX}) turned to unity at $r\rightarrow0$ and $\infty$.

The DMF of the diabatic covalent term is
\begin{eqnarray}
d^\textrm{d}_B(r)
&=&c_{0}+\sum_{i=1}^{6}c_{i}T_i\left(z_{1}\right)+\sum_{i=1}^{6}d_{i}T_i\left(z_{2}\right), \\
z_{k} &=&\tanh \left( a_{k}r-\frac{b_{k}}{r}\right) ,\quad k=1,2,
\end{eqnarray}
where parameters are restricted by relations 
\begin{eqnarray}
c_{0}+c_{2}+c_{4}+c_{6}+d_{2}+d_{4}+d_{6}=0, \\
c_{1}+c_{3}+c_{5}=0,\hspace{12pt} d_{1}+d_{3}+d_{5}=0,
\end{eqnarray}
thereby providing the fulfilness of the boundary conditions $d^\textrm{d}_B(0)=d^\textrm{d}_B(\infty )=0$ and the fitting convergence.
Finally,
\begin{eqnarray}
d^\textrm{d}_{XB}(r) &=&\frac{a\left( 1-e^{-cr}\right) ^{7}}{\sqrt{\left(
r^{2}-b^{2}\right) ^{4}+w^{4}}}{\sum_{i=0}^{6}\delta _{i}T_i(\xi)}
 , \\
\xi  &=&\tanh (\kappa r).
\end{eqnarray}
Parameters $\delta_i$ are restricted by the conditions for the sum to be unity at zero and infinity.
For the off-diagonal DMF, it was impossible to construct a regular function, the above model has a branch point.

Parameters of the model functions presented above will be optimized with use of the available experimental and \ai\ information
for the potentials and with the \ai\ data for other functions. The energy units are \cm{}, the bond length is in bohr for the NACME and angstrom for other functions.

\section{The datasets used in the fitting and the fitting results}\label{data}

The fitting of the molecular functions specified in Sec. \ref{molfunc} was performed to the following experimental and theoretical data for two isotopomers $^7$Li$^{19}$F and $^6$Li$^{19}$F.

1. The experimental microwave $\Delta J=1$ transition frequencies in the $v=0$-3 vibrational states \cite{Wharton63,Pearson69,Lovas74}.

2. The experimental infrared transition frequencies in the fundamental band $\Delta v=1,v\le8$ \cite{Hedderich91,Maki83}.

3. The \ai\ data of the present study.

All these data are included in Supplementary material.

Also, there are data that were not included in the fit, namely, the equilibrium dipole moments 
\cite{Hebert68,Lovas74} 
and the dipole moments of the $J = 1$ state in the $v=0$-3 vibrational states \cite{Wharton63,Hebert68,Lovas74}.


The Schr\"odinger equation in the diabatic basis, Eq. (\ref{eqsDiab}), is solved by the sinc-DVR method \cite{Colbert92sincDVR}. The Hamiltonian matrix has two diagonal blocks, each containing the sinc-DVR matrix of the kinetic energy and the potential energy calculated at the grid points, $r_i$, namely, $U_1^\textrm{d}$ and $U_2^\textrm{d}$ at the diagonals of the first and second blocks, respectively; they also include the centrifugal energy, $U_J$, and the nonadiabatic term, $-\mathcal{N}^2/2$, at $r_i$. The off-diagonal blocks of the Hamiltonian matrix contain only the diabatic couplings, $V^\textrm{d}(r_i)$, on their diagonals.

The grid points belong to the interval from $r_\textrm{min}$ to $r_\textrm{max}$, such that the wavefunction is zero at both ends. The number of grid points, $n$, defines the dimension of the matrix, hence, the time of matrix diagonalization $(\sim n^3)$, \emph{i.e.} the time needed for the parameter optimization. For the reliable calculation of the energies and wavefunctions, the grid step must be sufficiently lesser than the distance between the nodes of the wavefunctions of the vibrational states under consideration. Under the standard application of the sinc-DVR method, when a uniform grid is used, large matrices arise requiring much computer time. However, making the change of variable, it is possible to apply a non-uniform grid (the so-called adaptive sinc-DVR method \cite{Meshkov08}), so that the grid is dense in the range of oscillations of the wavefunction and rare where the wavefunction changes slowlier or decays under the potential barrier. Thus, the adaptive method permits to greatly diminish $n$ and to decrease the computer time by orders of magnitude.

The available experimental data on the transition frequencies cover the maximum vibration quantum number $v=8$ and $J<70$. In all cases, the outer turning point is reached at $r<2.2$ \AA. We used a non-uniform grid of 100 points within the interval of 0.3-10 \AA. The mapping function, 
\begin{equation*}
    y=\frac{r^2 - r_0^2}{r^2 + r_0^2},
\end{equation*}
was introduced with $r_0 = 1.56$ \AA{} and with a uniform grid along $y$.
Such a choice well satisfied the requirement for the wavefunctions to vanish at the end points and provided, with a large reserve, for convergence of the calculation of the energy spectrum.

Figure \ref{fig:NACME_N} shows the \ai\ calculated and the fitted NACME of Eq. (\ref{nacme1}). Interestingly, a second avoided crossing at short $r$ is discovered; its effect was not investigated.

Another important feature is the branch point of the adiabatic potentials near the avoided crossing, $r_\textrm{c}$. After the parameters are optimized, we find that the branch point is shifted to the upper half-plane by 0.128 \AA. At this point, the wave functions also have the respective brach points, therefore, $\mathcal{N}$ must have there a pole (according to Eq. (\ref{nacme1}), square root times square root gives pole). Indeed, it has, and the pole is shifted up by nearly the same value, 0.123 \AA, which certifies the self-consistency of the model functions.

\begin{figure}[htbp]
    \centering
    \includegraphics[width=0.7\linewidth]{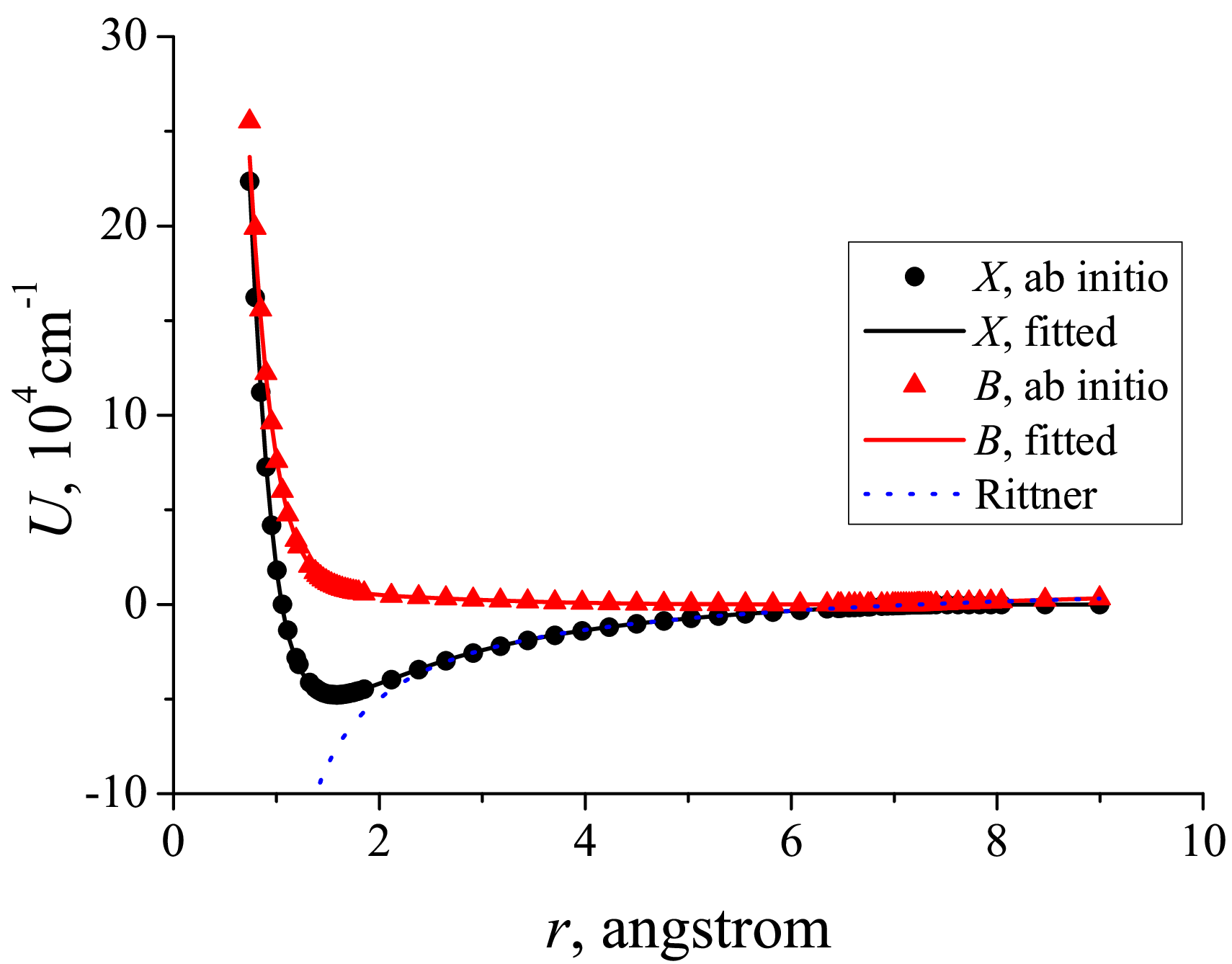}
    \includegraphics[width=0.7\linewidth]{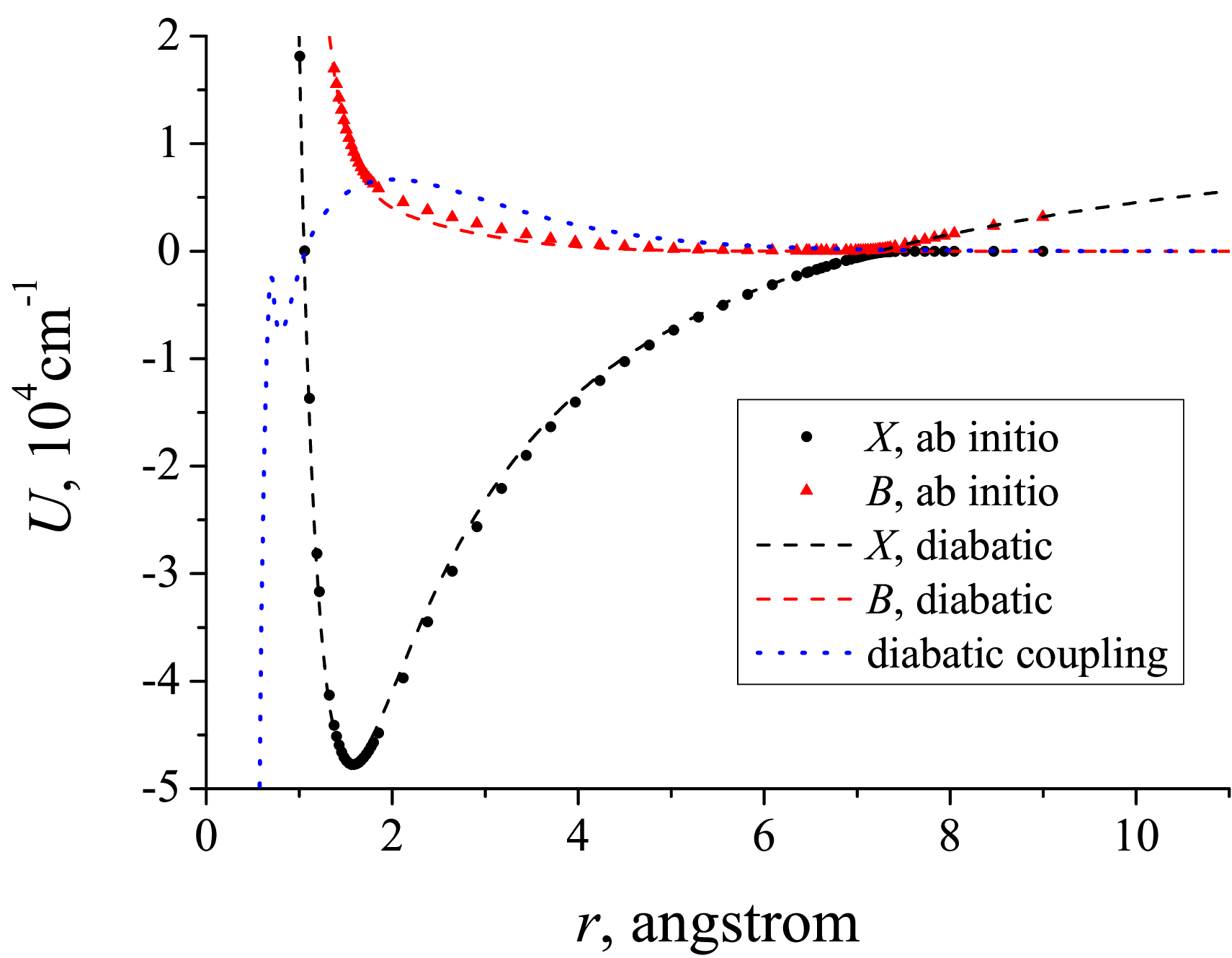}
    \caption{The BO adiabatic \ai\ (points) and fitted (full lines) potentials are shown along with the Rittner potential (dotted line) on the top panel; the BO diabatic potentials (dashed lines) and the diabatic coupling (dotted line) are shown on the bottom panel.}
    \label{fig:Uadiadia}
\end{figure}

\begin{figure}[htbp]
    \centering
    \includegraphics[width=0.7\linewidth]{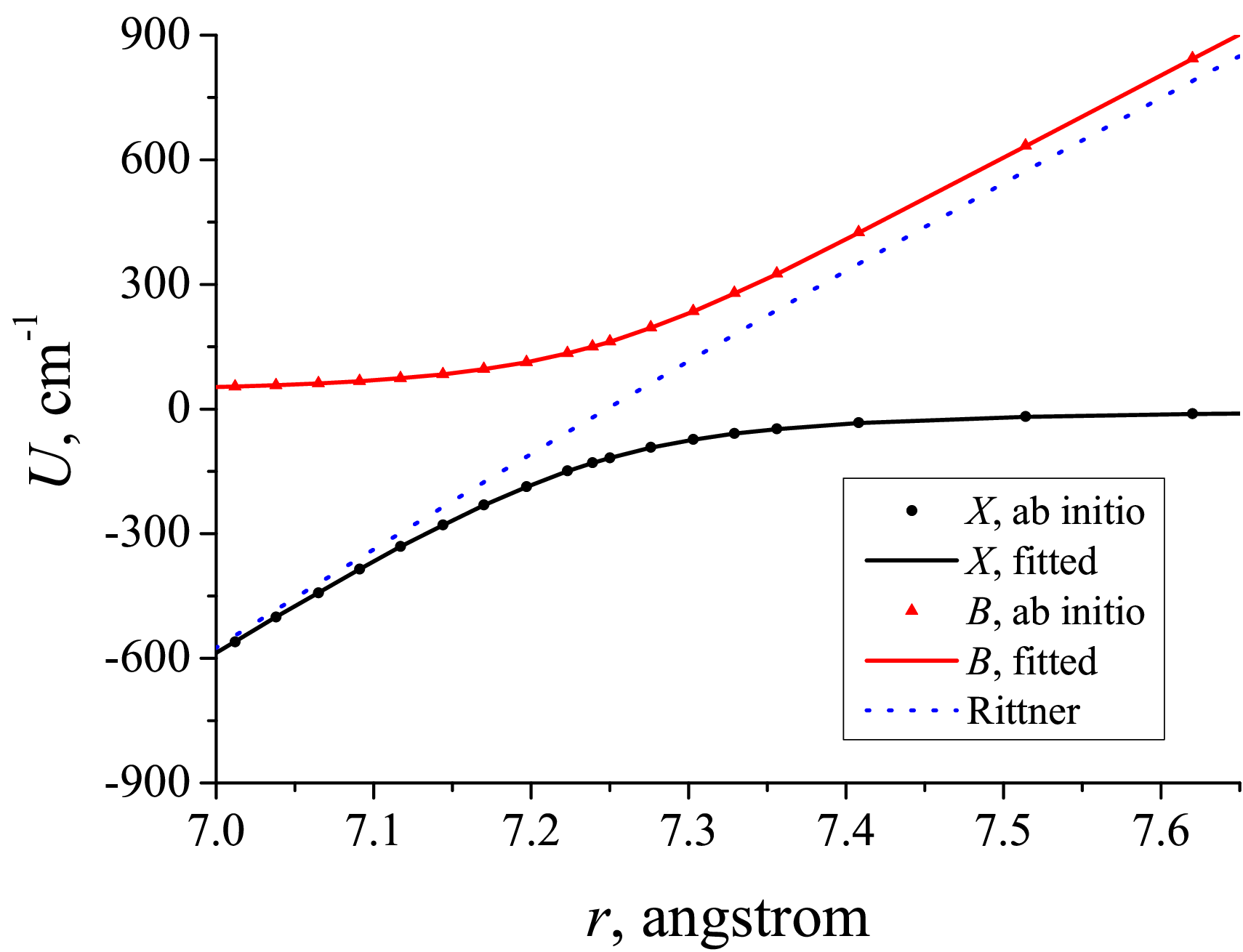}
    \includegraphics[width=0.7\linewidth]{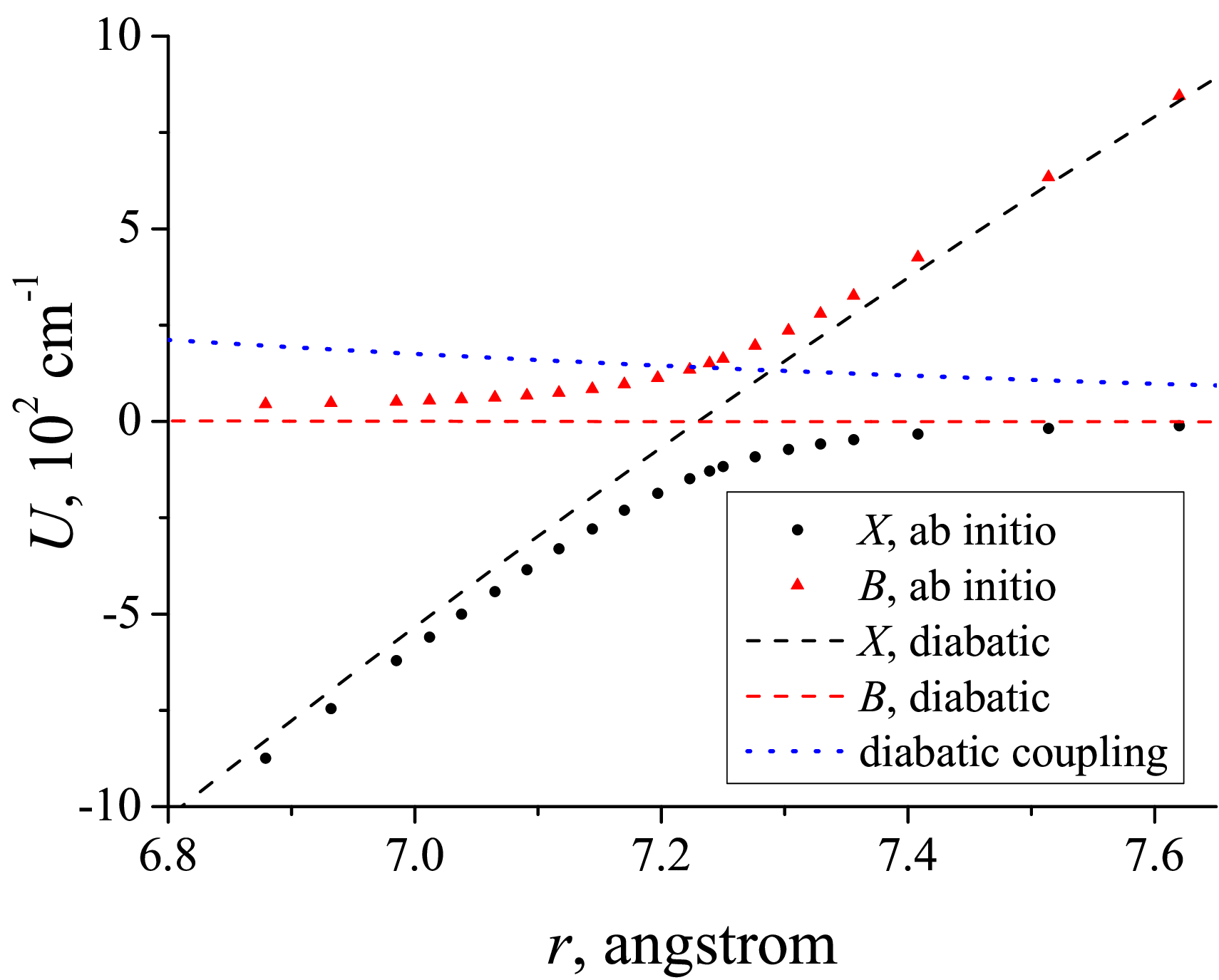}
    \caption{The same as in Fig. \ref{fig:Uadiadia} in the region of the avoided crossing.}
    \label{fig:Uadiadia_ac}
\end{figure}

The fitted adiabatic and diabatic potentials are shown in Figs. \ref{fig:Uadiadia} and \ref{fig:Uadiadia_ac}; Figure \ref{fig:Uadiadia} also shows the diabatic coupling. All diabatic functions are smooth everywhere including the vicinity of the avoided crossing, which greatly simplifies solution of two coupled equations.

It is seen in Figs. \ref{fig:Uadiadia} and \ref{fig:Uadiadia_ac} that the diabatic potential of the ionic state follows the Rittner potential very closely at $r>6$ bohr in accord with Refs. \cite{Kahn74,Bauschlicher88}. The diabatic coupling behave very smoothly in the crossing region, and, as seen from Fig. \ref{fig:Uadiadia_ac}, is nearly constant close to the minimum splitting between the adiabatic potentials, 139.7 cm$^{-1}$.

\begin{figure}[htbp]
    \centering
    \includegraphics[scale=0.4]{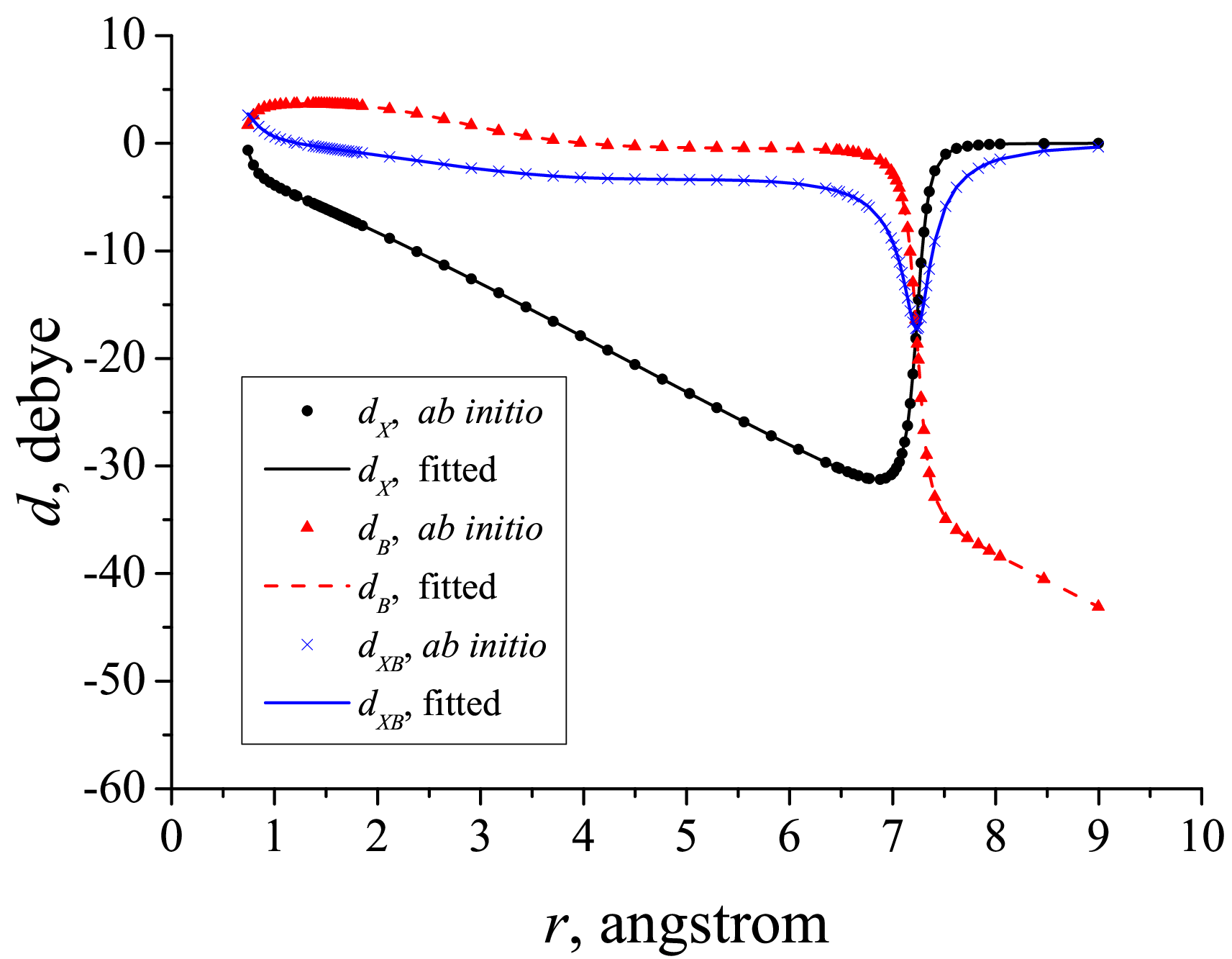}
    \caption{The \ai\ and fitted adiabatic dipole-moment functions.}
    \label{fig:DMFs}
\end{figure}

The fitted diagonal and off-diagonal DMFs and the fit quality are depicted in Fig. \ref{fig:DMFs}. The linear behavior of the DMF in the ionic $X$ state and its vanishing after the crossing region, and \emph{vice versa} for the neutral $B$ state are seen. The differences between the \ai\ and fitted DMFs do not exceed 0.02 D in the wide region.

\begin{figure}[htbp]
    \centering
    \includegraphics[scale=0.4]{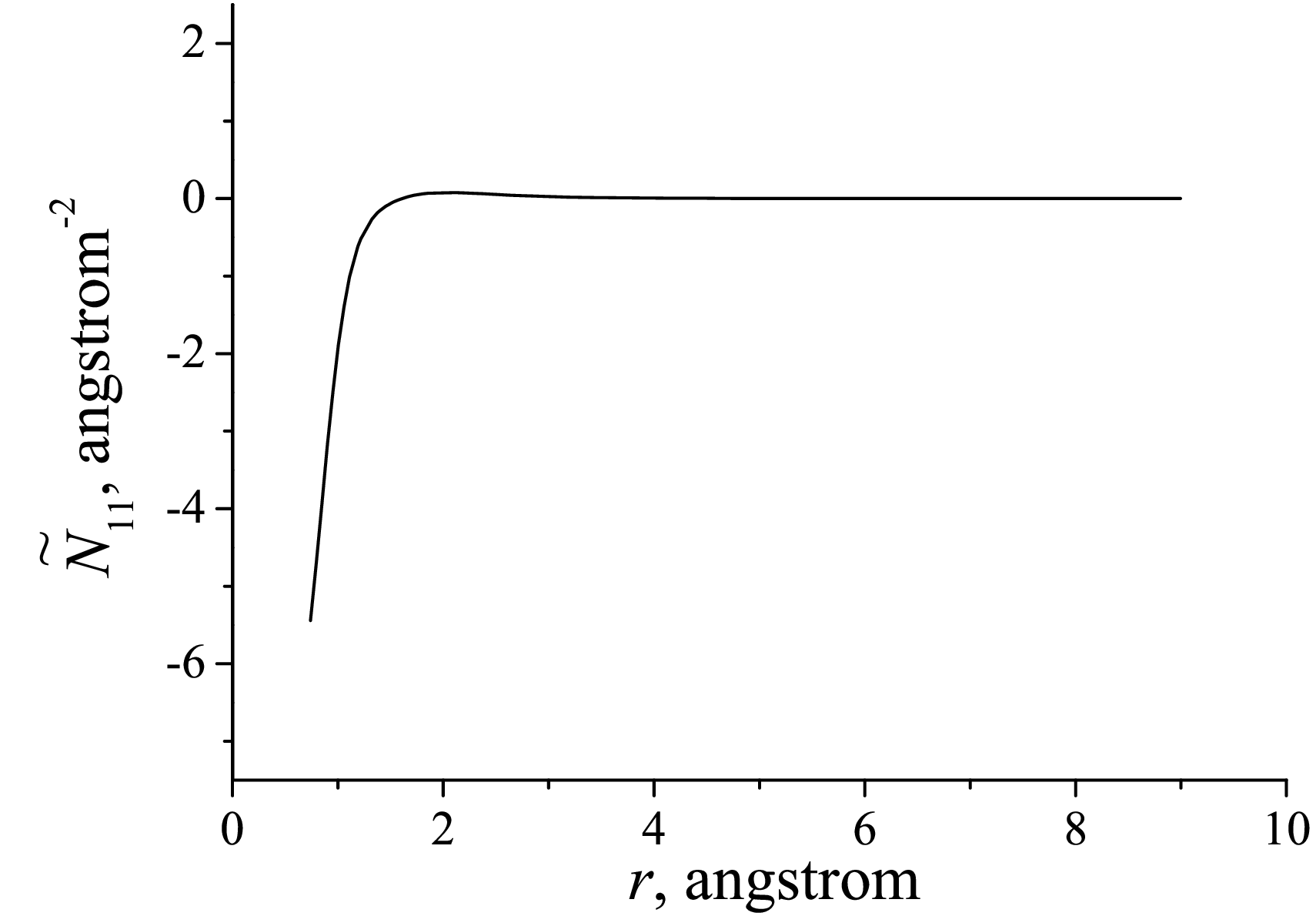}
    \caption{The mass-dependent nonadiabatic contribution to the ground-state potential. $\Tilde{N}_{11}=\mathcal{N}_{11}\cdot\mu/\hbar^2.$ }
    \label{fig:N11}
\end{figure}

\begin{figure}[htbp]
    \centering
    \includegraphics[scale=0.3]{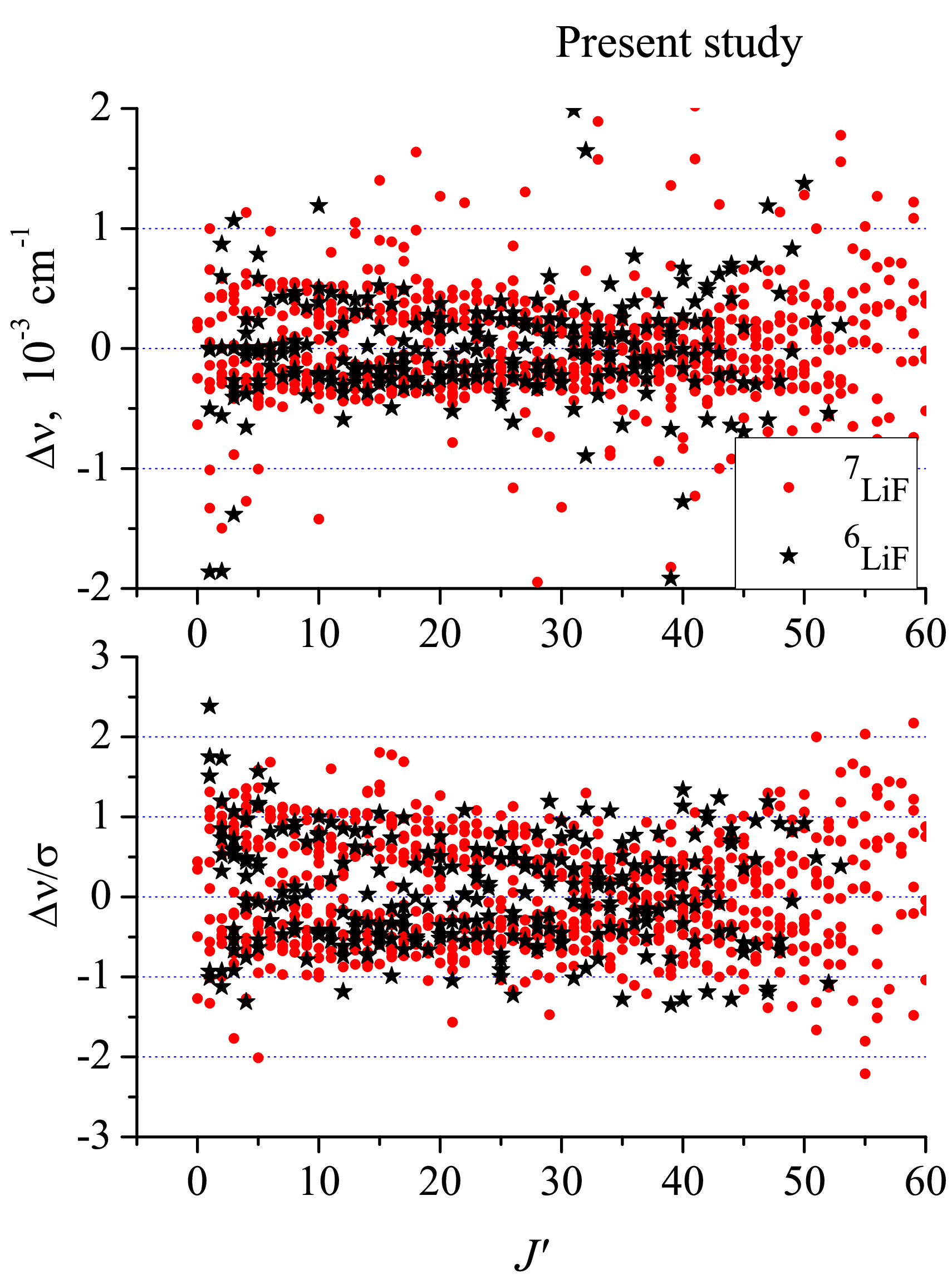}
    \caption{Observed minus calculated transition frequencies (top) and the same divided by the experimental error (bottom).}
    \label{fig:Delta_Nu_PS}
\end{figure}

The quality of reproduction of transition frequencies is demonstrated in Fig. \ref{fig:Delta_Nu_PS}. The differences are mostly within 0.0005 cm$^{-1}$ and, for the most of the lines, do not exceed the experimental uncertainty, $\sigma$; the reproduction quality is the same over all $J$ range, and the calculated frequencies do not show any apparent systematic deviations from the measured ones. Quantitatively, the root-mean-square deviations of the calculated frequencies from the measured ones is 0.0006 cm$^{-1}$ for $^6$LiF and 0.0009 cm$^{-1}$ for $^7$LiF, which agrees with the average values of the experimental errors, 0.00058 and 0.00076 cm$^{-1}$, respectively. For $^7$LiF, the scatter of the experimental errors is very large, from $5\cdot10^{-6}$ to 0.01 cm$^{-1}$; while the average error is 0.00076 cm$^{-1}$, the root-mean-square deviation is much larger, 0.0036 cm$^{-1}$. An important indicator of the fit quality is the reduced $\chi^2$, which is 0.5 for both $^6$LiF and $^7$LiF; the value appreciably less than unity probably certifies that $\sigma$ is overestimated.

\begin{figure}[htbp]
    \centering
    \includegraphics[scale=0.3]{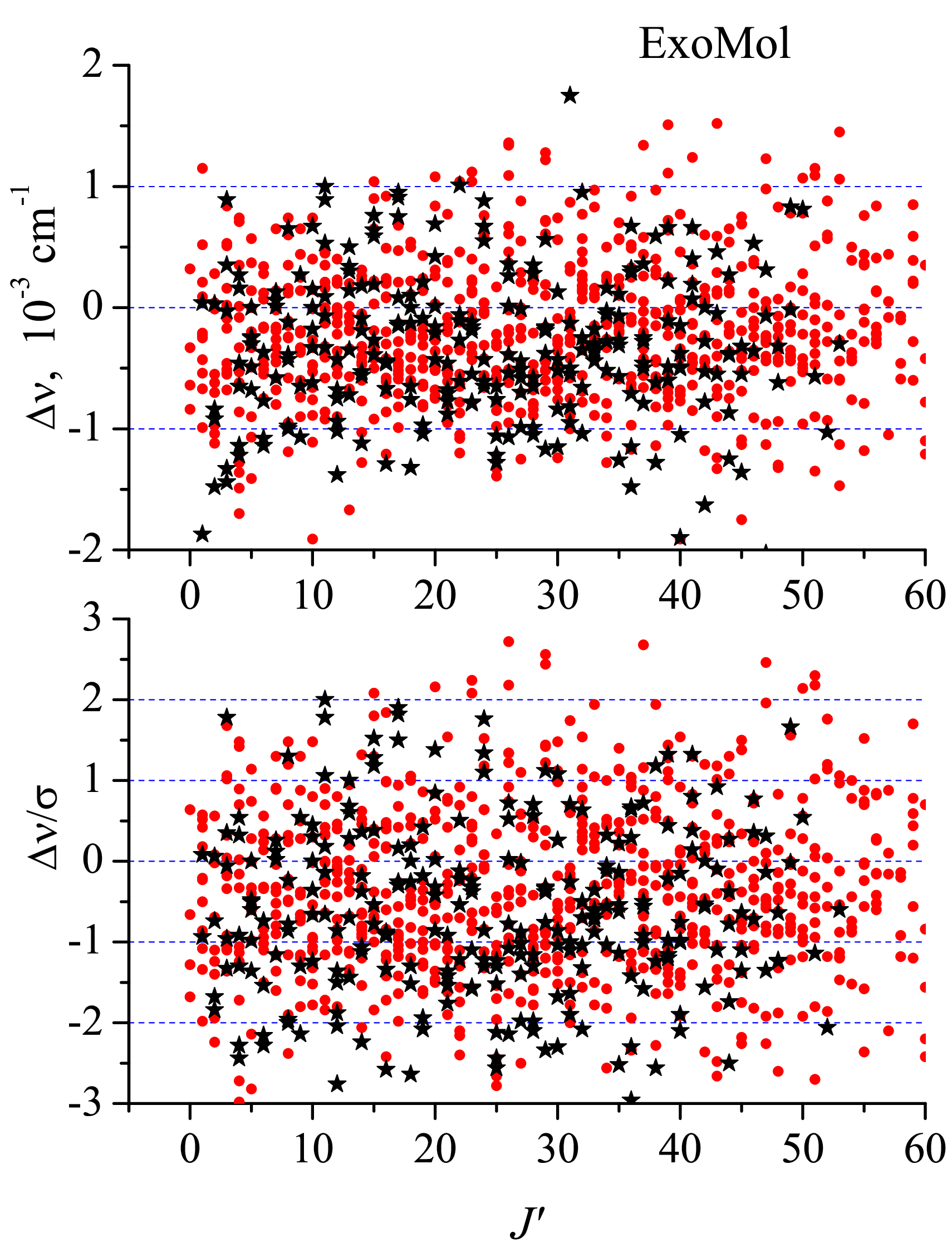}
    \caption{The same as in Fig. \ref{fig:Delta_Nu_PS} for the ExoMol data.}
    \label{fig:Delta_Nu_ExoMol}
\end{figure}

Figure \ref{fig:Delta_Nu_ExoMol} compares the ExoMol transition frequencies with experiment. The standard deviation\footnote{Calculated by us using the ExoMol frequencies published with three decimal places. \label{cbu}} of the calculated frequencies from the observed ones is only 0.001 \cm\, similar to the present study. The visible differences with the results of the present study shown in Fig. \ref{fig:Delta_Nu_PS} are presumably caused by truncation of the frequencies at the third decimal place, which is insufficient to correctly describe the high-precision observed data with $\sigma<10^{-3}$ \cm{}. 

\section{Resonances}\label{resonances}

When the energy is positive, the spectrum is continuous. 
Among the states found by solving the Schr\"odinger equation, there are such whose wavefunctions are localized predominantly at short $r$; they are responsible for the resonant scattering in the theory of gas-phase atomic collisions. Such resonant states can be of two kinds.

When $E$ is lower than the rotational barrier at a given high-enough $J$ ($J\ge17$), quasi-bound states, or resonances, with energy $E_J$ are formed whose widths, $\Gamma_J$, are due to the under-barrier tunneling.\footnote{At a given $J$, there are several resonances whose order number is omitted for brevity.} The quasi-bound level is subject to predissociation with the lifetime on the order of $\hbar/\Gamma_J$. Its width is mainly defined by the tunneling factor, yet the prefactor works as well because the top of the rotational barrier is situated in a vicinity of the avoided crossing where the $X$-$B$ interaction affecting the prefactor becomes strong. 

When $E$ is higher than the rotational barrier, the movement of two atoms becomes nearly diabatic, \emph{i.e.} they move in the diabatic potentials with a small probability to change the electronic state when moving through the region of the avoided crossing. 
If the interaction with the covalent state is neglected, the ionic diabatic state contains a set of discrete levels at $E>0$. Owing to the transitions, the system being initially in the ionic diabatic state may transfer to the repulsive covalent state with subsequent dissociation. In the theory of atomic collisions \cite{Nikitin84book,Voronin90book}, such states are called resonances, their width determines the probability of the resonant scattering. The widths of the resonances are defined not only by the above-mentioned transition probabilities, but also by the phase relations between the wavefunctions of the initial and final states in the region of the avoided crossing. It is important to emphasize that these relations may result in the resonance widths being much less than those due to the inter-state transitions alone. Namely, the waves may interfere in a vicinity of the crossing in such a way that the outgoing wave at infinity is nearly quenched. For instance, a $^7$LiF resonance exists whose width is extremely small, $\Gamma_J\approx4.6\cdot10^{-7}$ cm$^{-1}$. The transition probability itself behaves monotonously and varies relatively slowly, non-exponentially, with energy, whereas the widths of the nearby resonances vary chaotically due to the phase interference.

Finding the positions, $E_J$, and widths, $\Gamma_J$, of the quasi-bound states and resonances is performed by solving the Schr\"odinger equation in the diabatic basis, Eq. (\ref{eqsDiab}), using the sinc-DVR method \cite{Colbert92sincDVR}. The boundary conditions for the wave functions are to vanish at the end points of the grid, $r_\textrm{min}$ and $r_\textrm{max}$. The energies and wavefunctions thus defined depend on $r_\textrm{max}$, as illustrated in Figs. 5-8 of the paper by Mitev \etal\ for OH \cite{Mitev24_OH_prediss}. This can be used to find $E_J$'s and $\Gamma_J$'s.

In the flat Coulomb field of the ionic state, the wave functions of the high-lying bound states and resonances oscillate in a wide range of $r$, including the vicinity of the avoided crossing. In such a case, the DVR grid must fill in this range, therefore the application of the adaptive sinc-DVR method becomes inefficient. In calculations of the resonances and the bound-state rovibrational energy levels, we used the standard sinc-DVR method with a grid of 1200 points uniformly distributed within the interval from $r_\textrm{min}=0.3$ \AA{} to $r_\textrm{max}=15$ \AA{}.


Among solutions belonging to the continuous spectrum, there are those localized at short $r$ (the resonant solutions). At large $r$, they oscillate with small amplitudes, therefore their respective energies weakly depend on $r_\textrm{max}$. This permits to distinguish the resonant states from the non-resonant ones and to determine their characteristics.

At large $r$ and $E$ close to $E_J$, the phase of the adiabatic wavefunction of the ground electronic state at $r_\textrm{max}$ has the form\footnote{We remind that $E$ in this section stands for the energy of the discrete levels found by solution of Eq. (\ref{eqsDiab}) in a box; it depends on $J$, on the order number, and on the size of the box. However, to avoid confusion, we do not put any subscripts on $E$.} \cite{Landau77}
\begin{equation}\label{Phi}
    \Phi = \frac{1}{\hbar}S(E,r_\textrm{max})-\arctan\frac{\Gamma_J}{2\left(E-E_J\right)},
\end{equation}
where $S(E,r)$, the classical action, is the phase of the non-resonant scattering. In order to fulfill the boundary condition, $\Phi$ should not change with $r_\textrm{max}$,
\begin{equation}\label{Phiprime}
    \Phi^\prime=\frac{1}{\hbar}\frac{\partial S}{\partial E}E^\prime+k+\frac{2\Gamma_J}{4\left(E-E_J\right)^2+\Gamma_J^2}E^\prime = 0,
\end{equation} 
where prime stands for the derivative over $r_\textrm{max}$,
\begin{equation}\label{k}
k=\frac{1}{\hbar}\left.\frac{\partial S}{\partial
r}\right|_{r_\textrm{max}}=\frac{1}{\hbar}\sqrt{2\mu(E-U_\infty)},
\end{equation}
and $U_\infty$ is the ground-state potential limit at $r\rightarrow \infty$ (the dissociation limit). The partial derivative, $\partial S/\partial E$, approximately equals to the motion time from the barrier or from the avoided crossing to $r_\textrm{max}$,
\begin{equation}\label{dSdE}
\frac{\partial S}{\partial E} \sim
r_\textrm{max}\sqrt{\frac{\mu}{2(E-U_\infty)}}.
\end{equation}
For narrow resonances and not too large $r_\textrm{max}$, the first term
in Eq. (\ref{Phiprime}) is small, and we find from this equation 
\begin{equation}
E^\prime=-k\frac{4(E-E_J)^2+\Gamma_J^2}{2\Gamma_J}. \label{dEdR}
\end{equation}
For random values of $r_\textrm{max}$, the probability to obtain a certain value of $E$ is inversely proportional to $E^\prime$, and, as follows from Eq. (\ref{dEdR}), is given by the Lorentz distribution, \emph{cf.} Eq. (23) in Ref. \cite{Mitev24_OH_prediss}.

In order to determine the positions and widths of the resonances, Mitev \etal\ \cite{Mitev24_OH_prediss} calculated energies at various $r_\textrm{max}$ with subsequent statistical analysis of the results. This method requires much computer time since the matrix diagonalization is performed at each value of $r_\textrm{max}$; moreover, when the resonance is too narrow, the step in $r_\textrm{max}$ must be small, which makes the calculation unrealistic (which is also noted by Mitev \etal).

We explore the same idea of using the dependence of $E$ on $r_\textrm{max}$, but in a different way. Differentiating Eq. (\ref{dEdR}), we find the second derivative, 
\begin{equation}
E^{\prime \prime }= -\frac{k}{4\Gamma_J E}\left[4\left(E-E_J\right)^2+\Gamma_J^2\right]E^\prime -\frac{4k}{\Gamma_J}(E-E_{J})\,E^{\prime}.
\label{d2EdR}
\end{equation}
Using Eqs. (\ref{dEdR}) and (\ref{d2EdR}), we express the positions and widths of the resonances in terms of the energy derivatives over $r_\textrm{max}$,
\begin{equation}\label{E_J}
E_{J}=E-\delta E,
\end{equation}
where
\begin{equation}\label{delta E}
\delta E=\frac{4EE^{\prime ^{2}}\left( 2E^{\prime \prime
}E-E^{\prime ^{2}}\right) }{\left( 4kEE^{\prime }\right) ^{2}+\left(
2E^{\prime \prime }E-E^{\prime ^{2}}\right) ^{2}},
\end{equation}
\begin{equation}
\Gamma_J =-\frac{32kE^{2}E^{\prime ^{3}}}{\left( 4kEE^{\prime }\right)
^{2}+\left( 2E^{\prime \prime }E-E^{\prime ^{2}}\right) ^{2}}.
\label{Gamma}
\end{equation}
When $|E-E_J|\sim\Gamma_J$, the first term in Eq. (\ref{d2EdR}) is small; when, on the contrary, $|E-E_J|\ll\Gamma_J$, the first term is not small, but $\Gamma_J$ can be found directly from Eq. (\ref{dEdR}). After neglecting the first term in Eq. (\ref{d2EdR}), the equations acquire more compact forms,
\begin{equation}
E^{\prime \prime }=-\frac{4k}{\Gamma_J}(E-E_{J})\,E^{\prime }
\label{d2EdR0}
\end{equation}
\begin{eqnarray}
\delta E &=& \frac{2(E^{\prime })^{2}E^{\prime \prime }}{%
(E^{\prime \prime })^{2}+4k^{2}(E^{\prime })^{2}},  \label{deltaE0} \\
\Gamma_J &=&-\frac{8k(E^{\prime })^{3}}{(E^{\prime \prime
})^{2}+4k^{2}(E^{\prime })^{2}}. \label{Gamma0}
\end{eqnarray}
Calculations by Eqs. (\ref{deltaE0})-(\ref{Gamma0}) do not lead to any significant differences as compared to Eqs. (\ref{E_J})-(\ref{Gamma}).

After diagonalization of the Hamiltonian matrix and finding the eigenvalues and wavefunctions,
the energy derivatives can be calculated  
in the first (the Hellmann-Feynman theorem) and second orders of the perturbation theory, see Eqs. 
(94),
(96), and (103) in  Supplementary material.
Since all model functions are given in analytic forms, all necessary derivatives of the Hamiltonian matrix can be calculated analytically as well.

Far from resonances, or in their absence, the resonant second term in Eq. (\ref{Phiprime}) can be neglected. Then, we obtain
\begin{equation}
\frac{1}{\hbar }\frac{\partial S}{\partial E}E^{\prime }+k=0
\end{equation}
Inserting Eqs. (\ref{k}) and (\ref{dSdE}), we obtain for the uncoupled states
\begin{equation}\label{Eprime}
   E^\prime\approx -\frac{2(E-U_\infty)}{r_\textrm{max}}.
\end{equation}
The values of $E^\prime$ for bound or quasi-bound states are orders of magnitude smaller than those defined by this relation. Comparison of the energy derivative calculated after diagonalization of the Hamiltonian matrix with the one obtained by Eq. (\ref{Eprime}) enables us to efficiently filter out the unbound states; such a method has been earlier successfully applied by us to OH \cite{Ushakov25OH}.

In the case of the tunnel resonances, the widths of the resonances are exponentially decreasing with increasing the barrier width, therefore their calculation by Eqs. (\ref{delta E}) and (\ref{Gamma}) rapidly lose the accuracy because of the round-off errors. In this case, the quasi-classical expression can be used, 
\begin{equation}\label{tun_Gamma}
    \Gamma_J=\frac{\hbar\omega_J}{2\pi}\exp{\left(-D_J\right)},
\end{equation}
where $\omega_J$ is the vibration angular frequency in the potential well and $D_J$ is the tunneling exponent \cite{Landau77} in the effective adiabatic ground-state potential,
\begin{equation}\label{kappa}
    D_J = \frac{2}{\hbar}\int_{r_J^-}^{r_J^+}\sqrt{2\mu\left(U_1^\textrm{ad}+U_J-E_J\right)}dr,
\end{equation}
$r_J^-$ and $r_J^+$ being the inner and outer turning points at energy of the resonance, $E_J$.

In Eq. (\ref{tun_Gamma}), the interstate coupling is not taken into account. Comparison with calculations by Eqs. (\ref{E_J})-(\ref{Gamma}) shows that the above quasiclassical formula overestimates the width by about 10\%. The supplementary data files for the tunnel resonances contain both the results obtained with Eqs. (\ref{E_J})-(\ref{Gamma}) and with Eq. (\ref{tun_Gamma}). For $\Gamma_J<10^{-7}$ cm$^{-1}$, the formers become inaccurate, therefore we insert the quasiclassical values. 

The predissociation lifetimes of the resonances are given by the relation
\begin{equation}\label{tau}
    \tau_J=\frac{\hbar}{\Gamma_J}.
\end{equation}
We do not correct them for the above-mentioned inaccuracy of the quasiclassical formula. 

At $\delta E\ge\Gamma_J$, the accuracy of determination of the resonance positions and widths by Eqs. (\ref{E_J})-(\ref{Gamma}) decreases. In such cases, using Eqs. (\ref{Phi}) and (\ref{Phiprime}), we can calculate how $r_\textrm{max}$ should be changed (see Eq. 
(80)
in Supplementary material) in order that the new value of $\delta E$ became much smaller than $\Gamma_J$. In our calculations, this method is used to refine the parameters of all resonances with $\delta E/\Gamma_J\ge0.1$.

Additional notes about the search of the resonances and the calculations of the energy derivatives over $r_\textrm{max}$ are given in Supplementary material.

\section{Results and discussion}\label{Results}

\begin{table}[htbp]
    \centering
    \caption{Comparison of the molecular constants of the $X$ state calculated in the present study with experiment.}
    \vspace{7pt}
    \begin{tabular}{l|c|c|c|c|c|c|}
    \multicolumn{1}{c|}{} & \multicolumn{3}{|c|}{$^6$LiF}&\multicolumn{3}{|c|}{$^7$LiF}\\
    \hline
                                 &   PS      & Exp, ref. &  del,\%$^\dagger$ & PS    & Exp, ref. & del,\% \\
    \hline
      $r_\textrm{e}$, \AA{}      & 1.5644   & 1.5638 \cite{NIST} & 0.04 & 1.5644 & 1.5638 \cite{NIST} & 0.04 \\
      $\omega_\textrm{e}$, \cm\  & 965.6   & 964.3 \cite{Maki83} & 0.14 & 911.8  & 910.6 \cite{Maki83} & 0.13 \\
      $D_\textrm{e}$, \cm\       & & & & 48134 & 48393 \cite{Varandas09} & 0.54 \\
      $d_\textrm{eq}^\star$, D   & 6.40 & 6.28 \cite{Wharton63} & 1.9 & 6.40 & 6.28 \cite{Bittner18} & 1.9 \\
      $d_0$, D                   & 6.44 & 6.33 \cite{Hebert68} & 1.7 & 6.44 & 6.32 \cite{Hebert68} & 1.9 \\
      $d_1$, D                   & 6.53 & 6.41 \cite{Hebert68} & 1.9 & 6.52 & 6.41 \cite{Hebert68} & 1.7 \\
      $d_2$, D                   & 6.61 & 6.50 \cite{Hebert68} & 1.7 & 6.60 & 6.49 \cite{Hebert68} & 1.7\\
      $d_3$, D                   & 6.70 & 6.59 \cite{Wharton63} & 1.7 & & & \\
      \hline
      \multicolumn{7}{l}{$^\star d_\textrm{eq}=d_X^\textrm{ad}(r_\textrm{e}); d_v$, $v=0$-3, is given by Eq. (\ref{d_ad}) with $v^\prime=v^{\prime\prime}=v,J^{\prime\prime}=0$.}\\
      \multicolumn{7}{l}{$^\dagger100\left|\textrm{obs}-\textrm{calc}\right|/\textrm{obs}.$}
    \end{tabular}  
    \label{tab:molconst}
\end{table}

Table \ref{tab:molconst} shows some molecular constants calculated in the present study in comparison with the experimental data. The comparison shows that the potential of the present study is of high-enough quality since it correctly reproduces the measured molecular constants and can be used to calculate the energy levels and transition frequencies in LiF.

The line list calculated by Bittner and Bernath \cite{Bittner18} and used in the ExoMol project \cite{Wang20} is restricted by the vibrational quantum numbers of $v\le11$. 
This line list perfectly describes the data on transition frequencies, 
with the standard deviations (std) for both isotopologues$^{\ref{cbu}}$  
being only 0.001 cm$^{-1}$, and the same is true of the transition frequencies calculated in the present paper. Additionally, the quality of description of the experimental data can be characterized by the reduced chi-squared, $\chi_\textrm{red}=\sqrt{\chi^2/N_\textrm{points}}$, which is on the order of unity when the data are described within the experimental accuracy. Our calculations resulted in $\chi_\textrm{red}=0.5$ for both isotopologues, which certifies the excellent description (the experimental error being somewhat overestimated). For the ExoMol data, the same is true for the infrared data, yet $\chi_\textrm{red}$ cannot be calculated for the microwave data because the data were rounded off till the third decimal place only, which is insufficient for the high-precision (including the next three-four decimal places) experimental data.

One of the main results of the present study is the extension of the bound $X$ states up to the energies of 47600 cm$^{-1}$ close to the first dissociation limit where the presence of the excited state and the effect of the avoided crossing become appreciable. At such a high energy, the bound states disappear and the resonances appear instead. The calculated state energies are listed in Supplementary files \textbf{XLevels.dat} in zip-arxivs for two isotopologues.

\begin{table}[htbp]
    \centering
    \caption{Extract from supplementary file \textbf{Tun\_Resonances.dat} in \textbf{7LiF.zip} arxiv.$^\star$}
    \vspace{7pt}
    \begin{tabular}{|c|c|c|c|c|c|c|}
    \hline
    J  &  N     & EN$^a$       &     EN-E         & Gamma$^b$       & Tun\_Gamma$^c$ & tau/ps$^d$   \\
    \hline
   18  &  1     &    2.223489  &   47683.036492   &    5.446E-10    &   5.446E-10    &   9.748E+09   \\
   19  &  1     &    7.355656  &   47688.168660   &    4.109E-04    &   5.305E-04    &   1.292E+04   \\
   20  &  1     &   12.387971  &   47693.200975   &    3.972E-02    &   5.676E-02    &   1.337E+02   \\
   28  &  1     &    3.784460  &   47684.597464   &    8.275E-22    &   8.275E-22    &   6.415E+21   \\
\hline
\multicolumn{7}{l}{$^\star $Energies and widths in \cm, predissociation lifetime, $\tau$, in ps.}\\
\multicolumn{7}{l}{$^a$Eq. (\ref{delta E}); $^b$Eq. (\ref{Gamma}); $^c$Eq. (\ref{tun_Gamma}); $^d$Eq. (\ref{tau}).}
    \end{tabular}
    \label{tab:TunRes}
\end{table}

Table \ref{tab:TunRes} shows the tunneling resonances for $^7$LiF, its full version is given in Supplementary material along with the $^6$LiF data. The column headings are as follows: the rotation quantum number, $J$; the order number, $N$; the energy of the resonance, $E_J$, calculated by Eqs. (\ref{E_J}) and (\ref{delta E}), counted from the first dissociation limit; the same energy counted from the lowest bound level ($v=0,J=0$); the width of the resonance, $\Gamma_J$, calculated by Eqs. (\ref{Gamma}) and (\ref{tun_Gamma}); and the predissociation lifetime calculated by Eq. (\ref{tau}). For the most of the resonances, $\Gamma_J$ is calculated by Eq. (\ref{Gamma}); however, when $\Gamma_J$ becomes very small, $\Gamma_J<10^{-7}$ \cm\, the calculation becomes unreliable because of the round-off errors, in which case the value of col. ``Tun\_Gamma" is used, see the first and fourth rows in Table \ref{tab:TunRes}.

Supplementary files \textbf{Predissociation.dat} in two arxives, \textbf{$^7$LiF.zip} and \textbf{$^6$LiF.zip}, contain the resonances above the rotational barrier. Its column headings are the same as in \textbf{Tun\_Resonances.dat}, but the column ``Tun\_Gamma" is absent.

\begin{figure}[htbp]
    \centering
    \includegraphics[scale=0.4]{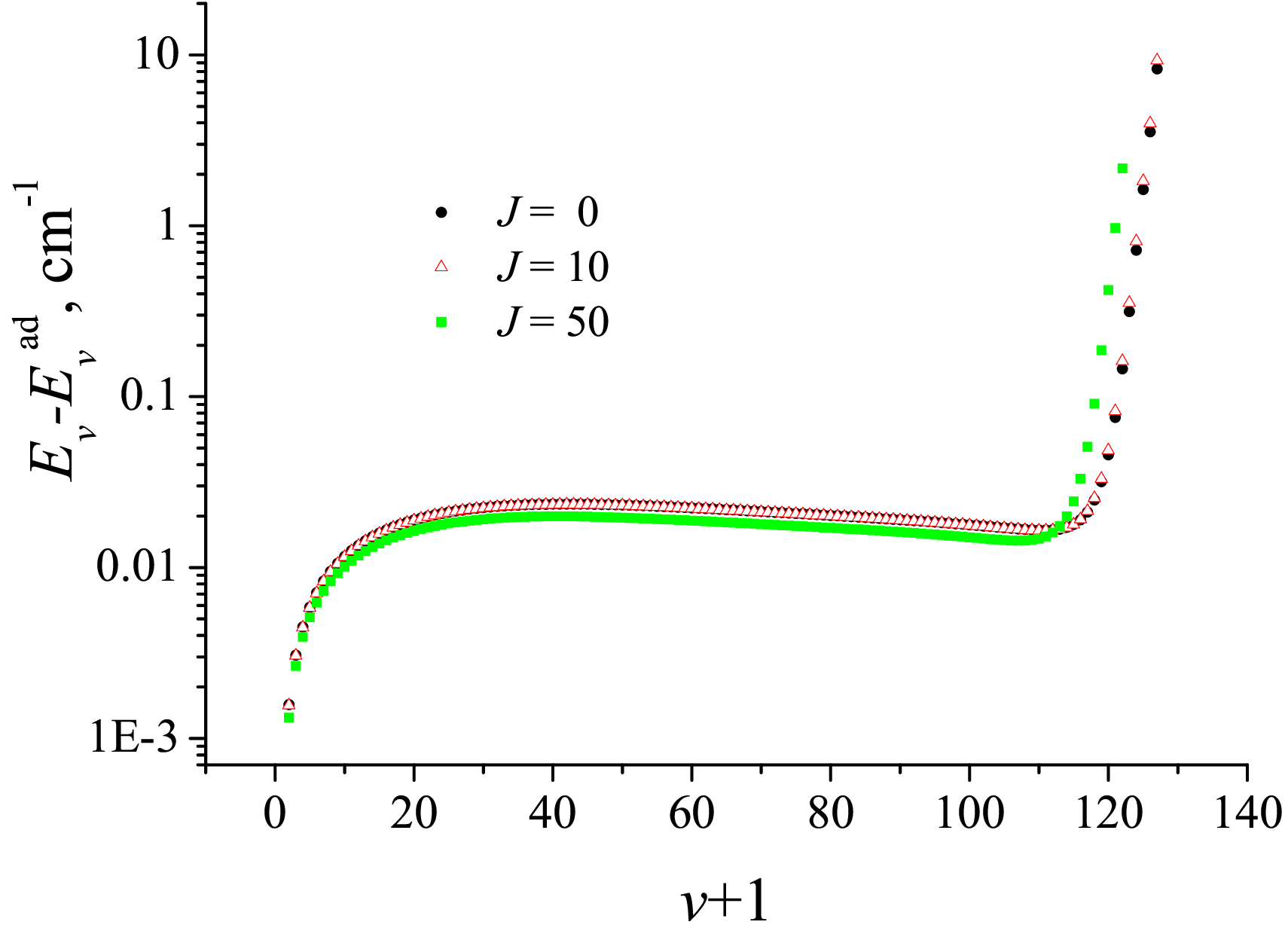}
    \caption{The changes in the level positions when NACME = 0. $E_v$, energies of the bound states; $E_v^\textrm{ad}$, the same energies calculated in the isolated ground-state adiabatic potential.}
    \label{fig:AdiaDia}
\end{figure}

Effect of the avoided crossing on the energies of the bound levels is shown in Fig. \ref{fig:AdiaDia}. Despite the smallness of the NACME, its effect on the level positions is significant owing to the high accuracy with which the transition frequencies are measured.

\begin{figure}[htbp]
    \centering
    \includegraphics[scale=0.4]{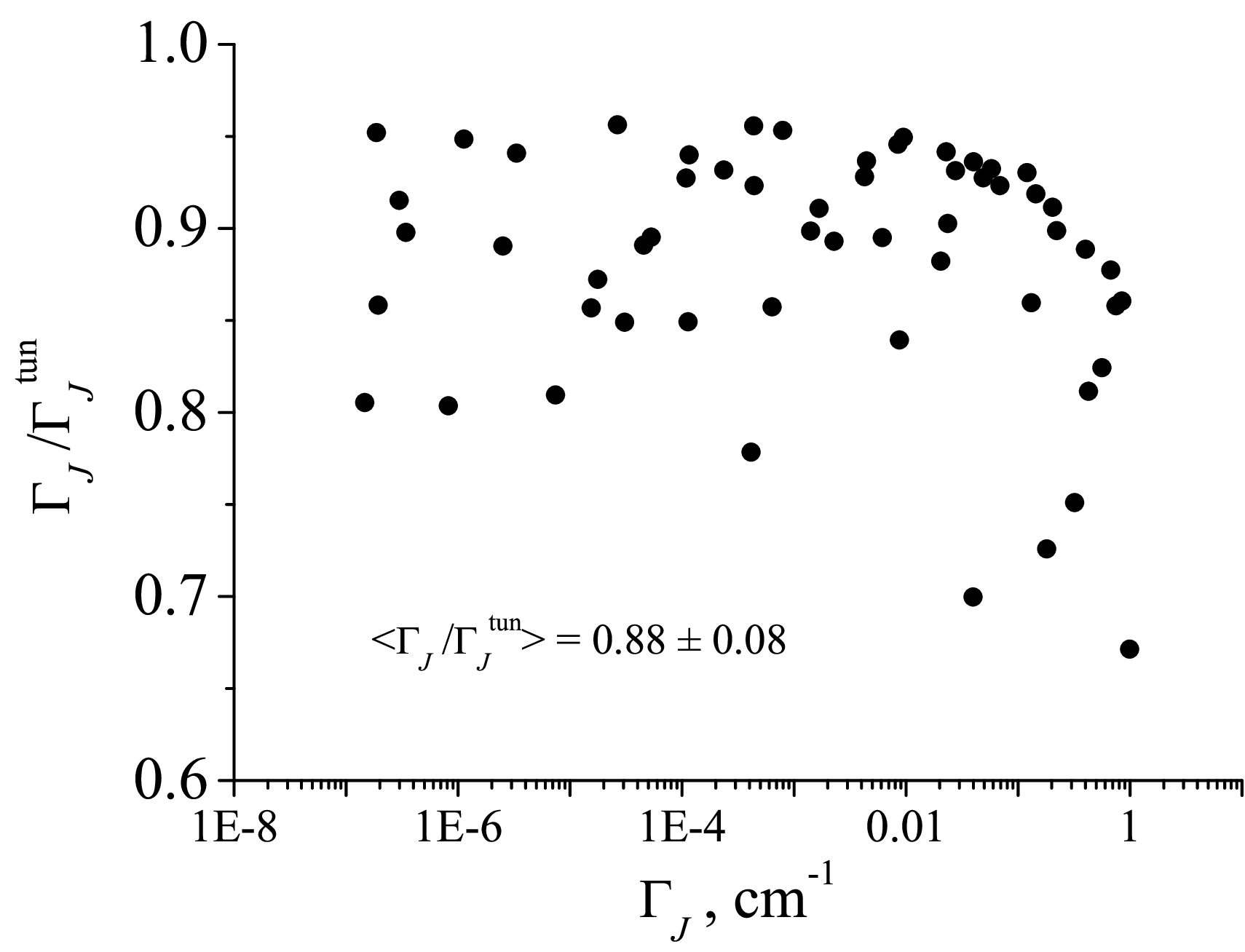}
    \caption{The ratio of the exact width (col. ``Gamma" in Table \ref{tab:TunRes}) to the resonance width at NACME = 0 calculated by the quasiclassical Eq. (\ref{tun_Gamma}) (col. ``Tun\_Gamma").}
    \label{fig:GGtun}
\end{figure}

The effect on the resonances is two-fold. When NACME = 0, the resonances above the rotational barrier turn into the bound states of the upper adiabatic term since no transitions between the ground and excited states can take place. The tunneling resonances occurring at energies below the barrier survive and become a bit wider, hence the lifetime shorter, as shown in Fig. \ref{fig:GGtun}. When NACME $\ne0$, the system in the ground adiabatic state may cross to the excited state, thereby increasing its time of residence at short $r$.

Finally, we calculated the line lists for two isotopologues including the bound-bound $X$-$X$ rovibrational transitions, $v\le50,\Delta v\le15,J\le170$ ($J\le200$ for the 0-0 and 1-0 bands); they are included in the Supplementary material. Figure \ref{fig:NIDL} shows the Einstein $A$ coefficients of the $v$-0R(0) lines in the NIDL \cite{Medvedev12} coordinates. For comparison, the ExoMol data for $v=0$-2 are shown. It is seen that the NIDL is perfectly followed for $v=2$-30. The difference with the ExoMol data is 2\% for $v=0$, 1\% for $v=1$, and 400\% for $v=2$. The failure to calculate the intensities of the $\Delta v\ge2$ transitions correctly, as described in Ref. \cite{Bittner18}, we explain by a small jump in the calculated \ai\ DMF near equilibrium.

\begin{figure}[htbp]
    \centering
    \includegraphics[scale=0.4]{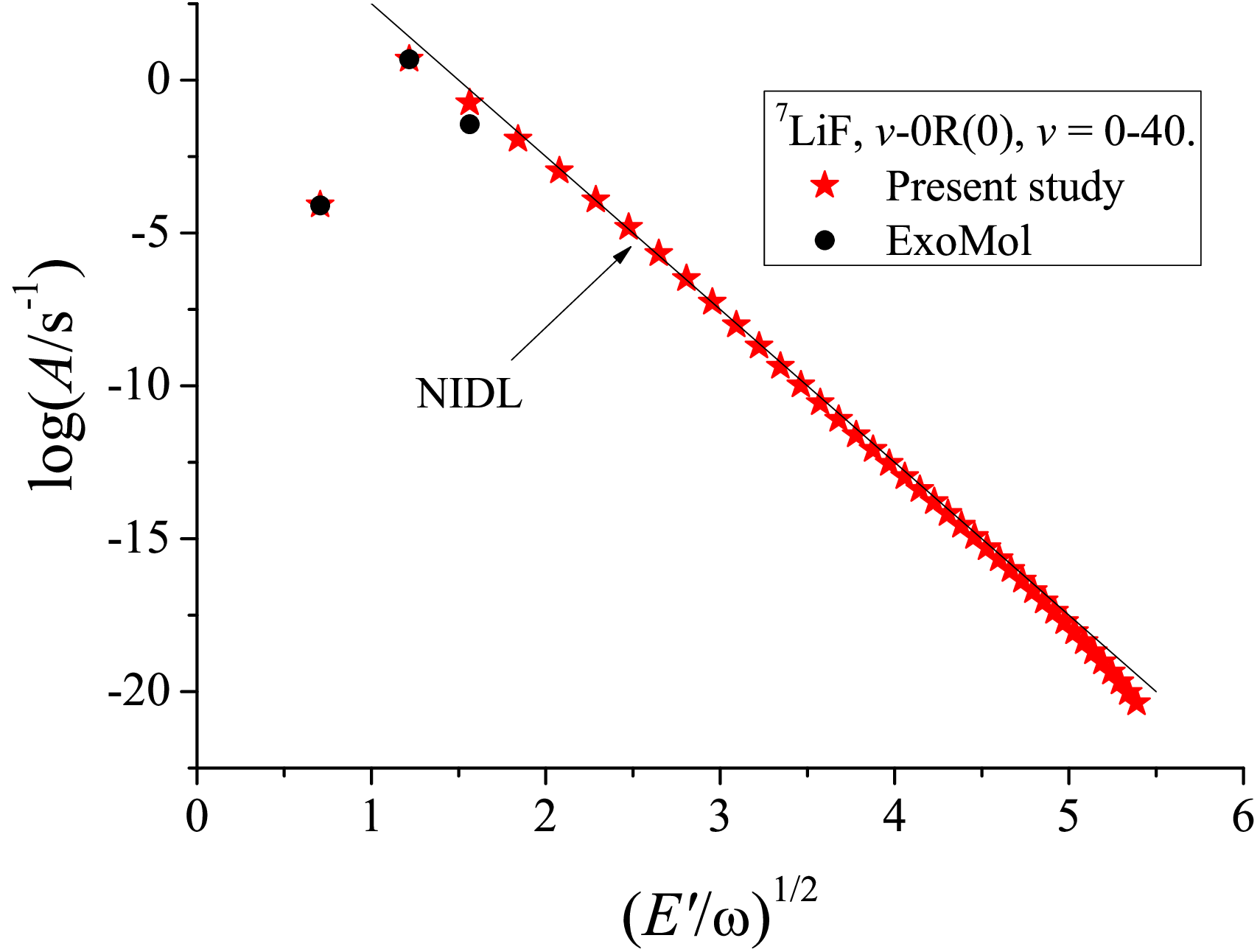}
    \caption{The calculated \EinA\ for the $v$-0R(0) lines in the NIDL \cite{Medvedev12} coordinates.}
    \label{fig:NIDL}
\end{figure}

\section{Conclusions}\label{Conclusions}

The system of two coupled Schr\"odinger equations is formulated, which includes the mass-dependent diagonal and off-diagonal nonadiabatic contributions of the static (\emph{i.e.} dependent on $r$ only) and dynamic (dependent on $r$ and velocity, the NACME) nature. All molecular functions are computed \ai\ and modeled with the analytic functions. The system is solved by the sinc-DVR method. Special attention is given to a new method of searching the resonances that has some advantages as compared to the one applied in Ref. \cite{Mitev24_OH_prediss} for OH. The theory is presented in full and can be applied to all alkali halides and other systems with the avoided crossings.  

The calculated energies of the bound states provide for excellent description of the observed line positions within the experimental uncertainties, as does the linelist of Bittner and Benath \cite{Bittner18} included in the ExoMol project \cite{Wang20}. The present study extends the calculated spectrum up to the dissociation limit.

The line lists for the rovibrational $X$-$X$ transitions in $^{6,7}$LiF are calculated. The \EinA\ perfectly follow the NIDL as predicted by the theory \cite{Medvedev12}. 
Calculation of the photodissociation cross sections and the absorption spectrum above the first dissociation limit is postponed to a future publication because it requires much additional work.

When the avoided crossing is accounted for (the NACME is not zero), the energies of states with $v\le110$ shift by small yet observable values, which rapidly increase at higher $v$. 

Two kinds of resonances are found at the energies above the first dissociation limit. When the energy is below the rotational barrier, the width of the resonance is mainly defined by the thickness of the barrier (the tunneling resonance). At higher energies, the resonances are all due to the possibility for the system to make transition at the avoided crossing and, by this reason, to be trapped for some time at short $r$. 

\section*{Acknowledgements}

We are grateful to P.~F.~Bernath for the data of Ref. \cite{Hedderich91} (which also included data of Refs. \cite{Wharton63,Pearson69}) and for permission to publish them. This work was performed under state tasks, Ministry of Science
and Education, Russian Federation, state registration numbers 124013000760-0 (FRC PCPMC) and 121031300176-3 (MSU).

\bibliography{Alcali_halides,ESMedvedev-Recovered-2,Overtones-Converted-Recovered-2}

\begin{thebibliography}{10}
\expandafter\ifx\csname url\endcsname\relax
  \def\url#1{\texttt{#1}}\fi
\expandafter\ifx\csname urlprefix\endcsname\relax\def\urlprefix{URL }\fi
\expandafter\ifx\csname href\endcsname\relax
  \def\href#1#2{#2} \def\path#1{#1}\fi

\bibitem{Grice74}
R.~Grice, D.~R. Herschbach, Long-range configuration interaction of ionic and covalent states, Mol. Phys. 27 (1974) 159--175.
\newblock \href {https://doi.org/10.1080/00268977400100131} {\path{doi:10.1080/00268977400100131}}.

\bibitem{Ewing71}
J.~J. Ewing, R.~Milstein, R.~S. Berry, Curve crossing in collisional dissociation of alkali halide molecules, J. Chem. Phys. 54 (1971) 1752--1760.
\newblock \href {https://doi.org/10.1063/1.1675082} {\path{doi:10.1063/1.1675082}}.

\bibitem{Kahn74}
L.~R. Kahn, P.~J. Hay, I.~Shavitt, {Theoretical study of curve crossing: \ai\ calculations on the four lowest $^1\Sigma^+$ states of LiF}, J. Chem. Phys. 61 (1974) 3530--3546.
\newblock \href {https://doi.org/10.1063/1.1682533} {\path{doi:10.1063/1.1682533}}.

\bibitem{Werner81}
H.-J. Werner, W.~Meyer, {MCSCF study of the avoided curve crossing of the two lowest $^1\Sigma^+$ states of LiF}, J. Chem. Phys. 74 (1981) 5802--5807.
\newblock \href {https://doi.org/10.1063/1.440893} {\path{doi:10.1063/1.440893}}.

\bibitem{Valiev20}
R.~R. Valiev, A.~A. Berezhnoy, I.~S. Gritsenko, B.~S. Merzlikin, V.~N. Cherepanov, T.~Kurten, C.~W\"ohler, Photolysis of diatomic molecules as a source of atoms in planetary exospheres, A\&{}A 633 (2020) A39.
\newblock \href {https://doi.org/10.1051/0004-6361/201936230} {\path{doi:10.1051/0004-6361/201936230}}.

\bibitem{Bittner18}
D.~M. Bittner, P.~Bernath, {Line lists for LiF and LiCl in the $X^1\Sigma^+$ ground state}, Astrophys. J. Suppl. Ser. 235 (2018) 8--12.
\newblock \href {https://doi.org/10.3847/1538-4365/aa9846} {\path{doi:10.3847/1538-4365/aa9846}}.

\bibitem{Bauschlicher88}
J.~Bauschlicher, C.~W., S.~R. Langhoff, {Full configuration‐interaction study of the ionic–neutral curve crossing in LiF}, J. Chem. Phys. 89 (1988) 4246--4254.
\newblock \href {https://doi.org/10.1063/1.455702} {\path{doi:10.1063/1.455702}}.

\bibitem{Giese04}
T.~J. Giese, D.~M. York, Complete basis set extrapolated potential energy, dipole, and polarizability surfaces of alkali halide ion-neutral weakly avoided crossings with and without applied electric fields, J. Chem. Phys., 120 (2004) 7939--7948.
\newblock \href {https://doi.org/10.1063/1.1690232} {\path{doi:10.1063/1.1690232}}.

\bibitem{Varandas09}
A.~J.~C. Varandas, {Accurate ab initio potential energy curves for the classic Li-F ionic-covalent interaction by extrapolation to the complete basis set limit and modeling of the radial nonadiabatic coupling}, J. Chem. Phys. 131 (2009).
\newblock \href {https://doi.org/10.1063/1.3237028. Erratum: J. Chem. Phys. 135 (2011) No. 11. doi: 10.1063/1.3641404} {\path{doi:10.1063/1.3237028. Erratum: J. Chem. Phys. 135 (2011) No. 11. doi: 10.1063/1.3641404}}.

\bibitem{Nkambule15}
S.~M. Nkambule, P.~Nurzia, A.~Larson, {Mutual neutralization in collisions of Li$^+$ and F$^-$}, Chem. Phys. 462 (2015) 23--27.
\newblock \href {https://doi.org/10.1016/j.chemphys.2015.08.006} {\path{doi:10.1016/j.chemphys.2015.08.006}}.

\bibitem{Smith69}
F.~T. Smith, {Diabatic and Adiabatic Representations for Atomic Collision Problems}, Phys. Rev. 179 (1969) 111--123.
\newblock \href {https://doi.org/10.1103/PhysRev.179.111} {\path{doi:10.1103/PhysRev.179.111}}.

\bibitem{Baer75}
M.~Baer, Adiabatic and diabatic representations for atom-molecule collisions: Treatment of the collinear arrangement, Chem. Phys. Lett. 35 (1975) 112--118.
\newblock \href {https://doi.org/10.1016/0009-2614(75)85599-0} {\path{doi:10.1016/0009-2614(75)85599-0}}.

\bibitem{Delos79}
J.~B. Delos, W.~R. Thorson, Diabatic and adiabatic representations for atomic collision processes, J. Chem. Phys. 70 (1979) 1774--1790.
\newblock \href {https://doi.org/10.1063/1.437650} {\path{doi:10.1063/1.437650}}.

\bibitem{Rittner51}
E.~S. Rittner, Binding energy and dipole moment of alkali halide molecules, J. Chem. Phys. 19 (1951) 1030--1035.
\newblock \href {https://doi.org/10.1063/1.1748448} {\path{doi:10.1063/1.1748448}}.

\bibitem{Child69}
M.~S. Child, Thermal energy scattering of alkali atoms from halogen atoms and molecules: the effect of curve crossing, Mol. Phys. 16 (1969) 313--327.
\newblock \href {https://doi.org/10.1080/00268976900100381} {\path{doi:10.1080/00268976900100381}}.

\bibitem{Voronin83}
A.~I. Voronin, V.~I. Osherov, L.~V. Poluyanov, V.~G. Ushakov, {Calculation of the cross sections of the recharge reactions H$_2^+$ (D$_2^+$) + Li = Li$^+$ + 2H (2D)}, Khimicheskaya Fizika 2 (1983) 28--34, in Russian.

\bibitem{Volokhov85}
V.~M. Volokhov, V.~I. Osherov, V.~G. Ushakov, {Calculation of the cross section of the dissociative recharge reaction H$_2^+$ + Mg}, Khimicheskaya Fizika 4 (1983) 1226--1230, in Russian.

\bibitem{MOLPRO2010}
H.-J. Werner, P.~J. Knowles, G.~Knizia, F.~R. Manby, M.~Sch\"utz, P.~Celani, T.~R. Korona, A.~Lindh, G.~Rauhut, K.~R. Shamasundar, T.~B. Adler, R.~D. Amos, A.~Bernhardsson, A.~Berning, D.~L. Cooper, M.~J.~O. Deegan, A.~J. Dobbyn, F.~Eckert, E.~Goll, C.~Hampel, A.~Hesselmann, G.~Hetzer, T.~Hrenar, G.~Jansen, C.~K\"oppl, Y.~Liu, A.~W. Lloyd, R.~A. Mata, A.~J. May, S.~J. McNicholas, W.~Meyer, M.~E. Mura, A.~Nicklass, D.~P. O'Neill, P.~Palmieri, K.~Pfl\"uger, R.~Pitzer, M.~Reiher, T.~Shiozaki, H.~Stoll, A.~J. Stone, R.~Tarroni, T.~Thorsteinsson, M.~Wang, A.~Wolf, \href{http://www.molpro.net.}{Molpro, version 2010.1, a package of ab initio programs} (2010).
\newline\urlprefix\url{http://www.molpro.net.}

\bibitem{Balakrishnan99}
N.~Balakrishnan, B.~D. Esry, H.~R. Sadeghpour, S.~T. Cornett, M.~J. Cavagnero, Quantum wave-packet dynamics of the photodissociation of {LiF}, Phys. Rev. A 60 (1999) 1407--1413.
\newblock \href {https://doi.org/10.1103/PhysRevA.60.1407} {\path{doi:10.1103/PhysRevA.60.1407}}.

\bibitem{Colbert92sincDVR}
D.~T. Colbert, W.~H. Miller, {A novel dicrete variable representation for quantum-mechanical reactive scattering via the S-matrix Kohn method}, J. Chem. Phys. 96 (1992) 1982--1991.
\newblock \href {https://doi.org/10.1063/1.462100} {\path{doi:10.1063/1.462100}}.

\bibitem{Meshkov08}
V.~V. Meshkov, A.~V. Stolyarov, R.~J. Le~Roy, {Adaptive analytical mapping procedure for efficiently solving the radial Schr\"odinger equation}, Phys. Rev. A 78 (2008) 052510.
\newblock \href {https://doi.org/10.1103/PhysRevA.78.052510} {\path{doi:10.1103/PhysRevA.78.052510}}.

\bibitem{Meshkov18}
V.~V. Meshkov, A.~V. Stolyarov, A.~Y. Ermilov, E.~S. Medvedev, V.~G. Ushakov, I.~E. Gordon, Semi-empirical ground-state potential of carbon monoxide with physical behavior in the limits of small and large inter-atomic separations, J. Quant. Spectrosc. Rad. Transfer 217 (2018) 262--273.
\newblock \href {https://doi.org/10.1016/j.jqsrt.2018.06.001} {\path{doi:10.1016/j.jqsrt.2018.06.001}}.

\bibitem{Wharton63}
L.~Wharton, W.~Klemperer, L.~P. Gold, R.~Strauch, J.~J. Gallagher, V.~E. Derr, Microwave spectrum, spectroscopic constants, and electric dipole moment of {Li$^6$F$^{19}$}, J. Chem. Phys. 38 (1963) 1203--1210.
\newblock \href {https://doi.org/10.1063/1.1733824} {\path{doi:10.1063/1.1733824}}.

\bibitem{Pearson69}
E.~F. Pearson, W.~Gordy, Millimeter- and submillimeter-wave spectra and molecular constants of {LiF and LiCl}, Phys. Rev. 177~(1) (1969) 52--58.
\newblock \href {https://doi.org/10.1103/PhysRev.177.52} {\path{doi:10.1103/PhysRev.177.52}}.

\bibitem{Lovas74}
F.~J. Lovas, E.~Tiemann, Microwave spectral tables {I. Diatomic} molecules, J. Phys. Chem. Ref. Data 3 (1974) 609--770.
\newblock \href {https://doi.org/10.1063/1.3253146} {\path{doi:10.1063/1.3253146}}.

\bibitem{Hedderich91}
H.~G. Hedderich, C.~I. Frum, R.~E. Jr., P.~F. Bernath, {The infrared emission spectra of LiF and HF}, Can. J. Chem. 69 (1991) 1659--1671.
\newblock \href {https://doi.org/10.1139/v91-244} {\path{doi:10.1139/v91-244}}.

\bibitem{Maki83}
A.~G. Maki, Infrared tunable diode laser spectra of lithium fluoride at high temperatures, J. Mol. Spectrosc. 102 (1983) 361--367.
\newblock \href {https://doi.org/10.1016/0022-2852(83)90046-2} {\path{doi:10.1016/0022-2852(83)90046-2}}.

\bibitem{Hebert68}
A.~J. Hebert, F.~J. Lovas, C.~A. Melendres, C.~D. Hollowell, J.~Story, T.~L., J.~Street, K., Dipole moments of some alkali halide molecules by the molecular beam electric resonance method, J. Chem. Phys. 48 (1968) 2824--2825.
\newblock \href {https://doi.org/10.1063/1.1669526} {\path{doi:10.1063/1.1669526}}.

\bibitem{Nikitin84book}
E.~E. Nikitin, S.~Y. Umanskii, Theory of Slow Atomic Collisions, Vol.~30 of Springer Series in Chemical Physics, Springer, Berlin, 1984.

\bibitem{Voronin90book}
A.~I. Voronin, Osherov, {Dynamics of Molecular Reactions (in Russian)}, Nauka, Moscow, 1990.

\bibitem{Mitev24_OH_prediss}
G.~B. Mitev, J.~Tennyson, S.~N. Yurchenko, {Predissociation dynamics of the hydroxyl radical (OH) based on a five-state spectroscopic model}, J. Chem. Phys. 160 (2024) 144110.
\newblock \href {https://doi.org/10.1063/5.0198241} {\path{doi:10.1063/5.0198241}}.

\bibitem{Landau77}
L.~D. Landau, E.~M. Lifshitz, Quantum Mechanics: Non-Relativistic Theory, 3rd Edition, Pergamon, Oxford, 1977.

\bibitem{Ushakov25OH}
V.~G. Ushakov, A.~Y. Ermilov, E.~S. Medvedev, Three-states model for calculating the $x$-$x$ rovibrational transition intensities in hydroxyl radical, J. Mol. Spectrosc. 407 (2025) 111977.
\newblock \href {https://doi.org/https://doi.org/10.1016/j.jms.2024.111977} {\path{doi:https://doi.org/10.1016/j.jms.2024.111977}}.

\bibitem{NIST}
 (2025).
\newblock \href{https://www.nist.gov/pml/molecular-spectroscopic-data}{[link]}.
\newline\urlprefix\url{https://www.nist.gov/pml/molecular-spectroscopic-data}

\bibitem{Wang20}
Y.~Wang, J.~Tennyson, S.~N. Yurchenko, {Empirical Line Lists in the ExoMol Database}, Atoms 8 (2020) 7.
\newblock \href {https://doi.org/10.3390/atoms8010007} {\path{doi:10.3390/atoms8010007}}.

\bibitem{Medvedev12}
E.~S. Medvedev, Towards understanding the nature of the intensities of overtone vibrational transitions, J. Chem. Phys. 137 (2012) 174307.
\newblock \href {https://doi.org/10.1063/1.4761930} {\path{doi:10.1063/1.4761930}}.

\end{thebibliography}
\bibliographystyle{elsarticle-num}

\end{document}